\documentclass[twocolumn]{aastex62}

\usepackage{natbib}
\usepackage{multirow}
\usepackage{listings}
\usepackage{color}
\usepackage{bm}
\usepackage{amsmath}

\shorttitle{RAPID}
\shortauthors{Muthukrishna et al.}

\begin{document}

\title{RAPID: Early Classification of Explosive Transients using Deep Learning}

\correspondingauthor{Daniel Muthukrishna}
\email{daniel.muthukrishna@ast.cam.ac.uk}

\author[0000-0002-5788-9280]{Daniel Muthukrishna}
\affiliation{Institute of Astronomy, University of Cambridge, Madingley
Road, Cambridge CB3 0HA, UK}

\author[0000-0001-6022-0484]{Gautham Narayan}
\altaffiliation{Lasker Fellow}
\affiliation{Space Telescope Science Institute, 3700 San Martin Dr., Baltimore, MD 21218, USA}

\author[0000-0001-9846-4417]{Kaisey S. Mandel}
\affiliation{Institute of Astronomy, University of Cambridge, Madingley
Road, Cambridge CB3 0HA, UK}
\affiliation{Statistical Laboratory, DPMMS, University of Cambridge, Wilberforce Road, Cambridge, CB3 0WB, UK}
\affiliation{Kavli Institute for Cosmology, Madingley Road, Cambridge, CB3 0HA, UK}

\author[0000-0002-5741-7195]{Rahul Biswas}
\affiliation{The Oskar Klein Centre for CosmoParticle Physics, Department of Physics, Stockholm University, AlbaNova, Stockholm SE-10691}

\author[0000-0002-0965-7864]{Ren\'{e}e Hlo\v{z}ek}
\affiliation{Department of Astronomy and Astrophysics \& Dunlap Institute, University of Toronto, 50 St. George Street, Toronto, ON M5S 3H4, CA}

\begin{abstract}
We present {\tt RAPID} (Real-time Automated Photometric IDentification), a novel time-series classification tool capable of automatically identifying transients from within a day of the initial alert, to the full lifetime of a light curve. Using a deep recurrent neural network with Gated Recurrent Units (GRUs), we present the first method specifically designed to provide early classifications of astronomical time-series data, typing 12 different transient classes. Our classifier can process light curves with any phase coverage, and it does not rely on deriving computationally expensive features from the data, making \texttt{RAPID} well-suited for processing the millions of alerts that ongoing and upcoming wide-field surveys such as the Zwicky Transient Facility (ZTF), and the Large Synoptic Survey Telescope (LSST) will produce. The classification accuracy improves over the lifetime of the transient as more photometric data becomes available, and across the 12 transient classes, we obtain an average area under the receiver operating characteristic curve of 0.95 and 0.98 at early and late epochs, respectively. We demonstrate \texttt{RAPID}'s ability to effectively provide early classifications of observed transients from the ZTF data stream. We have made \texttt{RAPID} available as an open-source software package\footnote{\url{https://astrorapid.readthedocs.io}} for machine learning-based alert-brokers to use for the autonomous and quick classification of several thousand light curves within a few seconds.
\end{abstract}


\section{Introduction} 
    \label{sec:Introduction}
    Observations of the transient universe have led to some of the most significant discoveries in astronomy and cosmology. From the use of Cepheids and type Ia supernovae (SNe Ia) as standardizable candles for estimating cosmological distances, to the recent detection of a kilonova event as the electromagnetic counterpart of GW170817, the transient sky continues to provide exciting new astronomical discoveries.
    
    In the past, transient science has had significant successes using visual classification by experienced astronomers to rank interesting new events and prioritize spectroscopic follow-up. Nevertheless, the visual classification process inevitably introduces latency into follow-up studies, and spectra for many objects are obtained several days to weeks after the initial detection. Existing and upcoming wide-field surveys and facilities will produce several million transient alerts per night, e.g. the Large Synoptic Survey Telescope \citep[LSST, ][]{Ivezic2009LSST:Products}, the Dark Energy Survey \citep[DES, ][]{DarkEnergySurveyCollaboration2016}, the Zwicky Transient Facility \citep[ZTF, ][]{Bellm2015TheFacility}, the Catalina Real-Time Transient Survey \citep[CRTS, ][]{Djorgovski2011TheCRTS}, the Panoramic Survey Telescope and Rapid Response System \citep[PanSTARRS, ][]{Chambers2018TheSurveys}, the Asteroid Terrestrial-impact Last Alert System \citep[ATLAS, ][]{Tonry2018ATLAS:System}, and the Planet Search Survey Telescope \citep[PSST, ][]{Dunham2004PSST:Telescope}. This unprecedented number means that it will be possible to obtain early-time observations of a large sample of transients, which in turn will enable detailed studies of their progenitor systems and a deeper understanding of their explosion physics. However, with this deluge of data comes new challenges, and individual visual classification for spectroscopic follow-up is utterly unfeasible. 
    
    Developing methods to automate the classification of photometric data is of particular importance to the transient community. In the case of SNe Ia, cosmological analyses to measure the equation of state of the dark energy $w$ and its evolution requires large samples with low contamination. The need for a high purity sample necessitates expensive spectroscopic observations to determine the type of each candidate, as classifying SNe Ia based on sparse light curves\footnote{We define light curves as photometric time-series measurements of a transient in multiple passbands. \textit{Full light curves} refer to time series of objects observed over nearly their full transient phase.  We refer to \textit{early light curves} as time series observed up to $2$ days after a trigger alert, defined in \S \ref{sec:trigger}.} runs the risk of contamination with other transients, particularly type Ibc supernovae. Even with human inspection, the differing cadence, observer frame passbands, photometric properties, and contextual information of each transient light curve constitute a complex mixture of sparse information, which can confound our visual sense, leading to potentially inconsistent classifications. This failing of visual classification, coupled with the large volumes of data, necessitates a streamlined automated classification process. This is our motivation for the development of our deep neural network (DNN) for Real-time Automated Photometric IDentification (\texttt{RAPID}), the focus of this work.

    \subsection{Previous Work on Automated Photometric Classification}
    
    In 2010, the Supernova Photometric Classification Challenge \citep[SNPhotCC,][]{snphotccChallenge,snphotccResults_SNII_1_SNIbc_1}, in preparation for the Dark Energy Survey (DES), spurred the development of several innovative classification techniques. The goal of the challenge was to determine which techniques could distinguish SNe Ia from several other classes of supernovae using light curves simulated with the properties of the DES. The techniques used for classification varied widely, from fitting light curves with a variety of templates \citep{Sako2008}, to much more complex methodologies that use semi-supervised learning approaches \citep{Richards2012} or parametric fitting of light curves \citep{Karpenka2012ANetworks}. A measure of the value of the SNPhotCC is that the dataset is still used as the reference standard to benchmark contemporary supernova light curve classification schemes, such as \citet{Bloom2011AutomatingEra,Richards2012,Ishida2013,Charnock2016,Lochner2016,Revsbech2017STACCATO:Sets,Narayan2018MachineStream,PELICANPasquet2019}. 
    
    Nearly all approaches to automated classification developed using the SNPhotCC dataset have either used empirical template-fitting methods \citep{Sako2008,Sako2011} or have extracted features from supernova light curves as inputs to machine learning classification algorithms \citep{Newling2011,Karpenka2012ANetworks,Lochner2016,Narayan2018MachineStream,Moller2016,Sooknunan2018}. \citet{Lochner2016} used a feature-based approach, computing features using either parametric fits to the light curve, template fitting with SALT2 \citep{Guy2007SALT2:Indicators}, or model-independent wavelet decomposition of the data. These features were independently fed into a range of machine learning architectures including Naive Bayes, $k$-nearest neighbours, multilayer perceptrons, support vector machines, and boosted decision trees (see \citealt{Lochner2016} for a brief description of these) and were used to classify just three broad supernova types. The work concluded that the non-parametric feature extraction approaches were most effective for all classifiers, and that boosted decision trees performed most effectively. Surprisingly, they further showed that the classifiers did not improve with the addition of redshift information. These previous approaches share two characteristics:
    \begin{enumerate}
        \item they are largely tuned to discriminate between different classes of supernovae, and
        \item they require the full phase coverage of each light curve for classification.
    \end{enumerate}
    Both characteristics arise from the SNPhotCC dataset. As it was developed to test photometric classification for an experiment using SNe Ia as cosmological probes, the training set represented only a few types of non-SNe Ia that were likely contaminants, whereas the transient sky is a menagerie. Additionally, SNPhotCC presented astronomers with full light curves, rather than the streaming data that is generated by real-time transient searches, such as ZTF. While previous methods can be extended with a larger, more diverse training set, the second characteristic they share is a more severe limitation. Requiring complete phase coverage of each light curve for classification (e.g. \citealt{Lochner2016}) is a fundamental design choice when developing the architecture for automated photometric classification, and methods cannot trivially be re-engineered to work with sparse data.
    
    \newpage
    \subsection{Early Classification}
    
    While retrospective classification after the full light curve of an event has been observed is useful, it also limits the scientific questions that can be answered about these events, many of which exhibit interesting physics at early-times. Detailed observations, including high-cadence photometry, time-resolved spectroscopy, and spectropolarimetry, shortly after the explosion provides insights into the progenitor systems that power the event and hence improves our understanding of the objects' physical mechanism. Therefore, ensuring a short latency between a transient detection and its follow-up is an important scientific challenge. Thus, a key goal of our work on \texttt{RAPID} has been to develop a classifier capable of identifying transient types within 2 days of detection. We refer to photometric typing with observations obtained in this narrow time-range as early classification. 
    
    The discovery of the electromagnetic counterpart \citep{LIGOScientificCollaboration2017Multi-messengerMerger,CoulterLIGO2017,Arcavi2017LIGO,SoaresSantosLIGO2017} from the binary neutron star merger event, GW170817, has thrown the need for automated photometric classifiers capable of identifying exotic events from sparse data into sharp relief. As we enter the era of multi-messenger astrophysics, it will become evermore important to decrease the latency between the detection and follow-up of transients. While the massive effort to optically follow up GW170817 was heroic, it involved a disarray of resource coordination. With the large volumes of interesting and unique data expected by surveys such as LSST ($\sim10^7$ alerts per night), the need to streamline follow-up processes is crucial. The automated early classification scheme developed in this work alongside the new transient brokers\footnote{Transient brokers are automated software systems that manage the real-time alert streams from transient surveys such as ZTF and LSST. They sift through, characterize, annotate and prioritise events for follow-up.} such as ALeRCE\footnote{\url{http://alerce.science}}, LASAIR\footnote{\url{https://lasair.roe.ac.uk}}~\citep{LASAIR}, ANTARES\footnote{\url{https://antares.noao.edu/}}~\citep{ANTARES, Saha16} are necessary to ensure organized and streamlined follow-up of the high density of exciting transients in upcoming surveys.
    
    There have been a few notable efforts at early-time photometric classification, particularly using additional contextual data. \citet{Sullivan2006PhotometricCandidates} successfully discriminated between SNe Ia and core-collapse SNe in the Supernova Legacy Survey using a template fitting technique on only two to three epochs of multiband photometry. \citet{Poznanski2007SingleEpoch} similarly attempted to distinguish between SNe Ia and core-collapse SNe using only single-epoch photometry along with a photometric redshift estimate from the probable host-galaxy. A few contemporary techniques such as \citet{Foley2013ClassifyingData} and the \texttt{sherlock} package\footnote{\url{https://github.com/thespacedoctor/sherlock}} use only host galaxy and contextual information with limited accuracy to predict transient classes (e.g. the metallicity of the host galaxy is correlated with supernova type). 
    
    The most widely used scheme for classification is \texttt{pSNid}~\citep{Sako2008,Sako2011}. It has been used by the Sloan Digital Sky Survey and DES \citep{DAndrea2018DES} to classify pre-maximum and full light curves into $3$ supernova types (SNIa, SNII, SNIbc). For each class, it has a library of template light curves generated over a grid of parameters (redshift, dust extinction, time of maximum, and light curve shape). To classify a transient, it performs an exhaustive search over the templates of all classes. It identifies the class of the template that best matches (with minimum $\chi^2$) the data and computes the Bayesian evidence by marginalizing the likelihood over the parameter space. The latest version employs  computationally-expensive nested sampling to compute the evidence. Therefore, the main computational burden (which increases with the number of templates used) is incurred every time it used to predict the class of each new transient. As new data arrives, this cost is multiplied as the procedure must be repeated each time the classifications are updated.
    
    In contrast, \texttt{RAPID} covers a much broader variety of transient classes and learns a function that directly maps the observed photometric time series onto these transient class probabilities. The main computational cost is incurred only once, during the training of the DNN, while predictions obtained by running new photometric time series through the trained network are very fast. Because of this advantage, updating the class probabilities as new data arrives is trivial. Furthermore, we specifically designed our RNN architecture for temporal processes.  In principle, it is able to save the information from previous nights so that the additional cost to update the classifications as new data are observed is only incremental.  These aspects make \texttt{RAPID} particularly well-suited to the large volume of transients that new surveys such as LSST will observe.
    
    \subsection{Deep Learning in Time-Domain Astronomy}
    
    Improving on previous feature-based classification schemes, and developing methods that exhibit good performance even with sparse data, requires new machine learning architectures. Advanced neural network architectures are non-feature-based approaches that have recently been shown to have several benefits such as low computational cost, and being robust against some of the biases that can afflict machine learning techniques that require ``expert-designed'' features \citep{Aguirre2017,Charnock2016,Moss2018,Naul2018AStars}. The use of Artificial Neural Networks \citep[ANN, ][]{McCulloch1943} and deep learning, in particular, has seen dramatic success in image classification, speech recognition, and computer vision, outperforming previous approaches in many benchmark challenges \citep{Krizhevsky2012,Razavian2014,Szegedy2015}.

    In time-domain astronomy, deep learning has recently been used in a variety of classification problems including variable stars \citep{Naul2018AStars,Hinners2017MachineClassification}, supernova spectra \citep{MuthukrishnaDASH}, photometric supernovae \citep{Charnock2016,Moss2018,SupernnoovaMoller2019,PELICANPasquet2019}, and sequences of transient images \citep{RCNN_Carrasco-Davis2018}. A particular class of ANNs known as Recurrent Neural Networks (RNNs) are particularly suited to learning sequential information (e.g. time-series data, speech recognition, and natural language problems). While ANNs are often feed-forward (e.g. convolutional neural networks and multilayer perceptrons), where information passes through the layers once, RNNs allow for cycling of information through the layers. They are able to encode an internal representation of previous epochs in time-series data, which along with real-time data, can be used for classification. 
    
    A variant of RNNs known as Long Short Term Memory Networks \citep[LSTMs, ][]{LSTM} improve upon standard RNNs by being able to store long-term information, and have achieved state-of-the-art performance in several time-series applications. In particular, they revolutionized speech recognition, outperforming traditional models \citep{Fernandez:2007:ARN:1778066.1778092, Hannun2014, Li2014} and have very recently been used in the trigger word detection algorithms popularized by \textit{Apple's Siri}, \textit{Microsoft's Cortana}, \textit{Google's voice assistant}, and \textit{Amazon's Echo}. \citet{Naul2018AStars} and \citet{Hinners2017MachineClassification} have had excellent success in variable star classification. \citet{Charnock2016} applied the technique to supernova classification. They used supernova data from the SNPhotCC and fed the multiband photometric full lightcurves into their LSTM architecture to achieve high SNIa vs non-SNIa binary classification accuracies. \citet{Moss2018} recently followed this up on the same data with a novel approach applying a new phased-LSTM \citep{pLSTM} architecture. These approaches have the advantage over previous supernova photometric classifiers of not requiring computationally-expensive and user-defined (and hence, possibly biased) feature engineering processes. 

    While this manuscript was under review, \citet{SupernnoovaMoller2019} released a similar algorithm for photometric classification of a range of supernova types. It uses a recurrent neural network architecture based on BRNNs (Bayesian Recurrent Neural Networks) and does not require any feature extraction. At a similar time, \citet{PELICANPasquet2019} released a package for full light curve photometric classification based on Convolutional Neural Networks, again able to use photometric light curves without requiring feature engineering processes. These approaches made use of datasets adapted from the SNPhotCC, using light-curves based on the observing properties of the Dark Energy Survey and with a smaller variety of transient classes than the PLAsTiCC-based training set used in our work. \citet{PELICANPasquet2019} use a framework that is very effective for full light curve classification, but is not well suited to early or partial light curve classification. \citet{SupernnoovaMoller2019}, on the other hand, is one of the first approaches able to classify partial supernova light curves using a single bi-directional RNN layer, and achieve accuracies of up to $96.92 \pm 0.26\%$ for a binary SNIa vs non-SNIa classifier. While their approach is similar, the type of RNN, neural network architecture, dataset, and focus on supernovae differs from the work presented in this paper. \texttt{RAPID} is focused on early and real-time light curve classification of a wide range of transient classes to identify interesting objects early, rather than on full light curve classification for creating pure photometric sample for SNIa cosmology. We are currently applying our work to the real-time alert stream from the ongoing ZTF survey through the ANTARES broker, and plan to develop the technique further for use on the LSST alert stream.
    
    \subsection{Overview}
    
    \begin{figure*}
	\centering
	\includegraphics[width=1.0\linewidth]{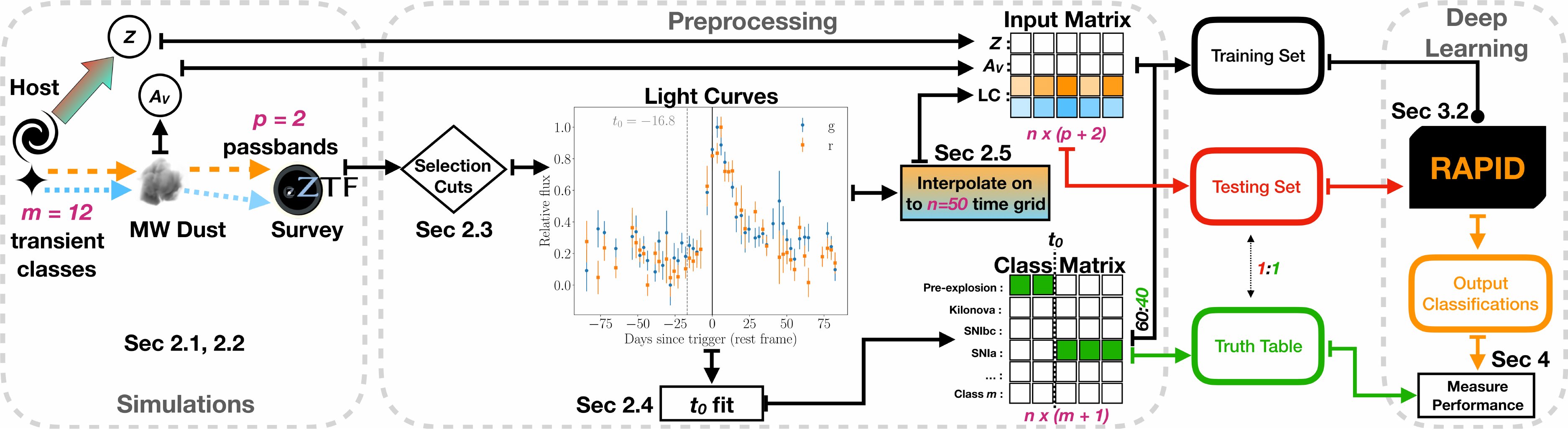}
	\caption{Schematic illustrating the preprocessing, training set preparation, and classification processes described throughout this paper.}
	\label{fig:matrix_schematic}
	\end{figure*}

    In this paper, we build upon the approach used in \citet{Charnock2016}. We develop \texttt{RAPID} using a deep neural network (DNN) architecture that employs a very recently improved RNN variation known as Gated Recurrent Units \citep[GRUs, ][]{GRU}. \textbf{This novel architecture allows us to provide real-time, rather than only full light curve, classifications.}
    
    Previous RNN approaches (including \citet{Charnock2016,Moss2018,Hinners2017MachineClassification,Naul2018AStars,SupernnoovaMoller2019}) all make use of bi-directional RNNs that can access input data from both past and future frames relative to the time at which the classification is desired.  While this is effective for full light curve classification, it does not suit the real-time, time-varying classification that we focus on in this work. In real-time classification, we can only access input data previous to the classification time. Therefore, to respect causality, we make use of uni-directional RNNs that only take inputs from time-steps previous to at any given classification time. \texttt{RAPID} also enables multi-class and multi-passband classifications of transients as well as a new and independent measure of transient explosion dates. We further make use of a new light curve simulation software developed by the recent Photometric LSST Astronomical Time-series Classification Challenge \citep[PLAsTiCC, ][]{plasticcNote}. 
    
   In section \ref{sec:Data} we discuss how we use the PLAsTiCC models with the SNANA software suite \citep{Kessler2010SNANA:Analysis} to simulate photometric light curves based on the observing characteristics of the ZTF survey, and describe the resulting dataset along with our cuts, processing, and modelling methods. In section \ref{sec:Model}, we frame the problem we aim to solve and detail the deep learning architecture used for \texttt{RAPID}. In section \ref{sec:Performance}, we evaluate our classifier's performance with a range of metrics, and in section \ref{sec:application_to_real_data} we apply the classifier to observed data from the live ZTF data stream.  An illustration of the different sections of this paper and their connections is shown in Fig.~\ref{fig:matrix_schematic}. Finally, in section \ref{sec:RandomForestClassifier}, we compare \texttt{RAPID} to a feature-based classification technique we implemented using an advanced Random Forest classifier improved from \citet{Narayan2018MachineStream} and based on \citet{Lochner2016}. We present conclusions in section 6.

\section{Data} 
    \label{sec:Data}
    \subsection{Simulations}
    \label{sec:Simulations}
        One of the key challenges with developing classifiers for upcoming transient surveys is the lack of labelled samples that are appropriate for training. Moreover, even once a survey commences, it can take a significant amount of time to accumulate a well-labelled sample that is large enough to develop robust learning algorithms. To meet this difficulty for LSST, the PLAsTiCC collaboration has developed the infrastructure to simulate light curves of astrophysical sources with realistic sampling and noise properties. This effort was one component of an open-access challenge to develop algorithms that classify astronomical transients. By adapting supernova analysis tools such as SNANA \citep{Kessler2010SNANA:Analysis} to process several models of astrophysical phenomena from leading experts, a range of new transient behavior included in the PLAsTiCC dataset. The challenge has recently been released to the public on Kaggle\footnote{\url{https://www.kaggle.com}} \citep{plasticcNote} along with the metric framework to evaluate submissions to the challenge \citep{plasticcMetric}. The PLAsTiCC models are the most comprehensive enumeration of the transient and variable sky available at present.
        
        We use the PLAsTiCC transient class models and the simulation code developed in \citet{KesslerPlasticcModels} to create a simulated dataset that is representative of the cadence and observing properties of the ongoing public ``Mid Scale Innovations Program'' (MSIP) survey at the ZTF \citep{Bellm2015TheFacility}. This allows us to compare the validity of the simulations with the live ZTF data stream, and apply our classifier to it as illustrated in section \ref{sec:application_to_real_data}.
        
        \subsubsection{Zwicky Transient Facility}
        ZTF is the first of the new generation of optical synoptic survey telescopes and builds upon the infrastructure of the Palomar Transient Factory \citep[PTF, ][]{PTFrau2009}. It employs a 47 square degree field-of-view camera to scan more than 3750 square degrees an hour to a depth of 20.5 - 21 mag \citep{GrahamZTFProceedings}. It is a precursor to the LSST and will be the first survey to produce one million alerts a night and to have a trillion row data archive. To prepare for this unprecedented data volume, we build an automated classifier trained on a large simulated ZTF-like dataset that contains a labelled sample of transients. 
        
\begin{figure}
        	\centering
        	\includegraphics[width=1.\linewidth]{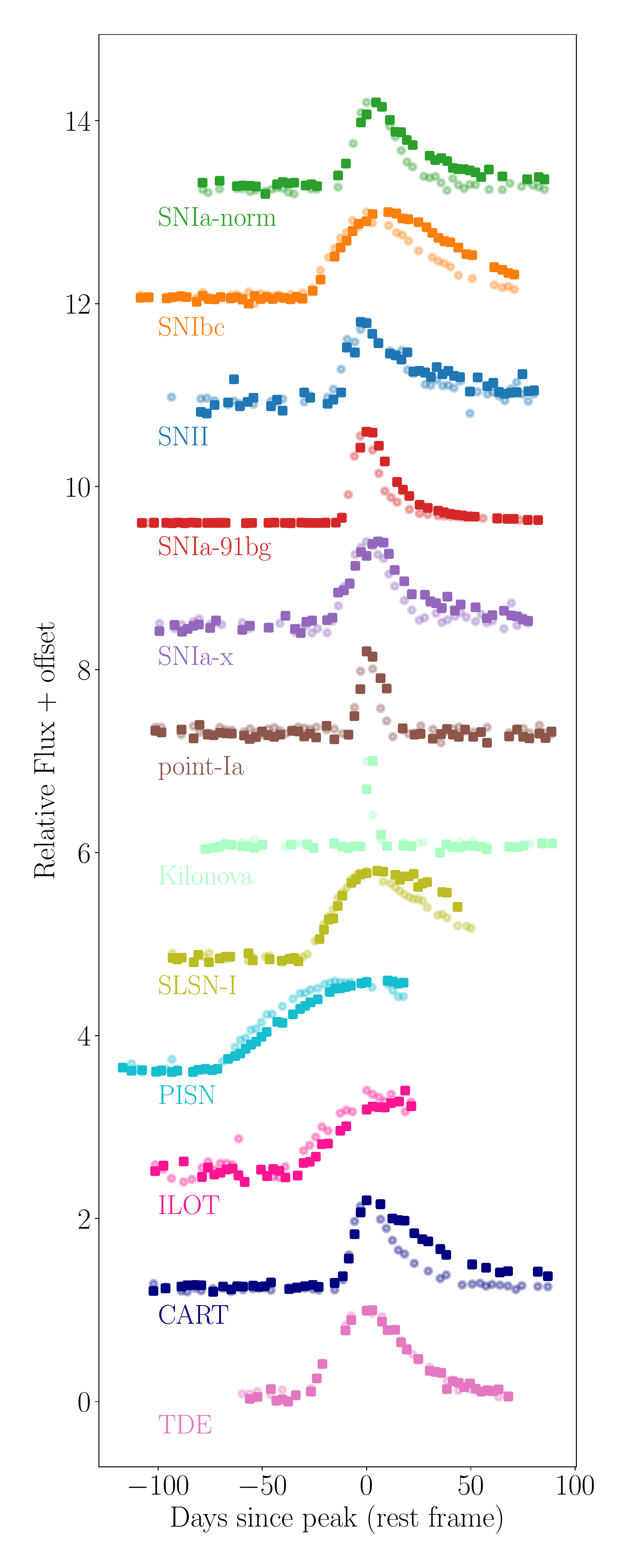}
        	\caption{The light curves of one example transient from each of the 12 transient classes is plotted with an offset. We have only plotted transients with a high signal-to-noise and with a low simulated host redshift ($z<0.2$) to facilitate comparison of light curve shape between the classes. The dark-coloured square markers plots the $r$ band light curves of each transient, while the lighter-coloured circle markers are the $g$ band light curves of each transient.}
        	\label{fig:example_classes_lc}
    	\end{figure}

        We obtained logs of ZTF observing conditions (E. Bellm, private communication) and photometric properties (zeropoints, FWHM, sky brightness etc.), and one of us (R.B.) converted these into a library suitable for use with SNANA. SNANA simulates millions of light curves for each model, following a class-specific luminosity function prescription within the ZTF footprint. The sampling and noise properties of each observation on each light curve is set to reflect a random sequence from within the observing conditions library. The simulated light curves thus mimic the ZTF observing properties with a median cadence of 3 days in the $g$ and $r$ passbands. As ZTF had only been operating for four months when we constructed the observing conditions library, it is likely that our simulations are not fully representative of the survey. Nevertheless, this procedure is more realistic than simulating the observing conditions entirely, as we would have been forced to do if we had developed \texttt{RAPID} for LSST or \emph{WFIRST}. We verified that the simulated light curves have similar properties to observed transient sources detected by ZTF that have been announced publicly. The dataset consists of a labelled set of $48029$ simulated transients evenly distributed across a range of different classes briefly described below. An example of a simulated light curve from each class is shown in Fig.~\ref{fig:example_classes_lc}.
        
        \subsubsection{Transient Classes}
        We briefly describe the transient class models from \citealt{KesslerPlasticcModels} that are used throughout this paper.
        
        \begin{description}
            \item[Type Ia Supernovae]
                Type Ia supernovae are the thermonuclear explosion of a binary star system consisting of a carbon-oxygen white dwarf accreting matter from a companion star. In recent years, many subgroups in the SNIa class have been defined to account for their observed diversity. In this work we include three subtypes. SNIa-norm are the most commonly observed SNIa class. Type Ia-91bg Supernovae (SNIa-91bg) burn at slightly lower luminosities and have lower eject velocities. SNIax are similar, and are SN2002cx-like supernovae \citep[defined in ][]{Silverman2012,Foley2013-Iax}.
            \item[Core collapse Supernovae]
                Type Ibc (SNIbc) and Type II (SNII) supernovae are typically found in regions of star formation, and are the result of the core collapse of massive stars. Their light curve shape and spectra near maximum light look very similar to SNIa, but tend to have magnitudes about 1.0-1.5 times fainter than a typical SNIa.
            \item[point-Ia]
                These are a hypothetical supernova type which are expected to have light curve shapes very similar to normal SNe Ia, but are just one-tenth as bright. They are the result of the early onset of detonation of helium transferring white dwarf binaries known as AM Canum Venacticorum systems \citep{pointIa_1}. Helium that accretes onto carbon-oxygen white dwarfs undergoes unstable thermonuclear flashes when the orbital period is short: in the 2.5--3.5 minute range  \citep{Bildsten2007FaintBinaries}. This process is strong enough to result in the onset of a detonation. 
            \item[TDE]
                Tidal Disruption Events occur when a star in the orbit of a massive black hole is pulled apart by the black hole's tidal forces. Some debris from the event is ejected at high speeds, while the remainder is swallowed by the black hole, resulting in a bright flare lasting up to a few years \citep{TDE_3}.
            \item[Kilonovae] 
                Kilonovae have been observed as the electromagnetic counterparts of gravitational waves. They are the mergers of either double neutron stars (NS-NS) or black hole neutron star (BH-NS) binaries, the former of which was recently discovered by LIGO \citep[see the famous GW170817][]{LIGOScientificCollaboration2017Multi-messengerMerger,Abbott2017GW170817:Inspiral}. The neutron-rich ejecta from the merger undergoes rapid neutron capture (r-process) nucleosynthesis to produce the Universe's rare heavy elements. The radioactive decay of these unstable nuclei power a rapidly evolving transient kilonova event \citep{Metzger2016Kilonovae,Yu2017AKilonova}.
            \item[SLSN-I] 
                Type I Super-luminous supernovae (SLSN) have $\sim 10$ times the energy of SNe Ia and core-collapse SNe, and are thought to be caused by several different progenitor mechanisms including magnetars, the core-collapse of particularly massive stars, and interaction with circum-stellar material. In analogy to supernovae, they are divided into hydrogen-poor (type I) and hydrogen-rich (type II). A subclass of SLSN-I appear to be powered by radioactive decay of Ni-56, and are termed SLSN-R. However, the majority of SLSN-I require some other energy source, as the nickel mass required to power their peaks is in conflict with their late-time decay. While SLSN-I and SLSN-II have different spectroscopic fingerprints, their light curves are qualitatively similar, and for classification purposes, it suffices to consider SLSN-I as a proxy for classification performance on all SLSN. See SN2005ap in \citet{Quimby2007SNExplosion} for an example and \citet{Galyam2018SLSNeReview} for a comprehensive review.
            \item[PISN]
                Pair-instability Supernovae are thought to be runaway thermonuclear explosions of massive stars with oxygen cores initiated when the internal energy in the core is sufficiently high to initiate pair production. This pair-production from $\gamma$ rays in turn leads to a dramatic drop in pressure support, and partial collapse. The rapid contraction leads to accelerated oxygen ignition, followed by explosion. These explosive transients can only result from stars with masses $\sim 10$~\(M_\odot\) and above, and they naturally yield several solar masses of Ni-56. \citet{Ren2012ASDSS} and \citet{PISN_SLSNR} has suggested that observed members of the SLSN-R subclass are consistent with PISN models. 
            \item[ILOT]
                Intermediate Luminosity Transients have a peak luminosity in the energy gap between novae and supernovae (e.g. NGC 300 OT2008-1, \citealt{Berger2009AnStar}). The physical mechanism of these objects is not well understood, but they have been modelled as either the eruption of red giants or as interacting binary systems (see \citealt{Kashi2016AnASASSN-15qi} and references therein). 
            \item[CART]
                Calcium-rich gap transients (e.g. PTF11kmb, \citealt{Lunnan2017TwoEnvironments}) are a recently discovered transient class that have strong forbidden and permitted calcium lines in their spectra. The physical mechanism of these events is not well understood, but they are known to evolve much faster than average SNe Ia with rise times less than 15 days (compared with $\sim18$ days for SNIa), have velocities of approximately $6000$ to $10000$ $\mathrm{km \ s^{-1}}$, and have absolute magnitudes in the range -15.5 to -16.5 (a factor of 10 to 30 times fainter than SNe Ia) \citep{Sell2015Calcium-RichDwarfs,Kasliwal2012CALCIUM-RICHGALAXIES_CART_3}.

        \end{description}

        The above list of transients is not exhaustive, but is the largest collection of transient models assembled to date. The specific models used in the simulations derived from \citet{KesslerPlasticcModels} are SNIa-norm: \citet{SNIa_1, SNIa_2, SNIa_3_SNII_2_SNIbc_2}, SNIbc: \citet{snphotccResults_SNII_1_SNIbc_1, SNIa_3_SNII_2_SNIbc_2, SNII_3_SNIbc_3_SLSN_1_TDE_1_ILOT_1_CART_1_PISN_1, SNII_4_SNIbc_4_ILOT_2_CART_2_PISN_2}, SNII: \citet{snphotccResults_SNII_1_SNIbc_1, SNIa_3_SNII_2_SNIbc_2, SNII_3_SNIbc_3_SLSN_1_TDE_1_ILOT_1_CART_1_PISN_1, SNII_4_SNIbc_4_ILOT_2_CART_2_PISN_2}, SNIa-91bg: (Galbany et al. in prep.), SNIa-x: \citet{SNIax_1}, pointIa: \citet{pointIa_1}, Kilonovae: \citet{KN_1}, SLSN: \citet{SNII_3_SNIbc_3_SLSN_1_TDE_1_ILOT_1_CART_1_PISN_1, SLSN_2, SLSN_3}, PISN: \citet{SNII_3_SNIbc_3_SLSN_1_TDE_1_ILOT_1_CART_1_PISN_1, SNII_4_SNIbc_4_ILOT_2_CART_2_PISN_2, PISN_3}, ILOT: \citet{SNII_3_SNIbc_3_SLSN_1_TDE_1_ILOT_1_CART_1_PISN_1, SNII_4_SNIbc_4_ILOT_2_CART_2_PISN_2}, CART: \citet{SNII_3_SNIbc_3_SLSN_1_TDE_1_ILOT_1_CART_1_PISN_1, SNII_4_SNIbc_4_ILOT_2_CART_2_PISN_2, Kasliwal2012CALCIUM-RICHGALAXIES_CART_3}, TDE: \citet{SNII_3_SNIbc_3_SLSN_1_TDE_1_ILOT_1_CART_1_PISN_1, TDE_2, TDE_3}. 
        
        Each simulated transient dataset consists of a time series of flux and flux error measurements in the $g$ and $r$ ZTF bands, along with sky position, Milky Way dust reddening, a host-galaxy redshift, and a photometric redshift. The models used in PLAsTiCC were extensively validated against real observations by several complementary techniques, as described by \citet[][in prep.]{NarayanPlasticcValidation}. We split the total set of transients into two parts: 60\% for the \textit{training set} and 40\% for the \textit{testing set}. The \textit{training set} is used to train the classifier to identify the correct transient class, while the \textit{testing set} is used to test the performance of the classifier.

        \subsection{Trigger for Issuing Alerts}\label{sec:trigger}
            The primary method used for detecting transient events is to subtract real-time or archival data from a new image to detect a change in observed flux. This is known as \textit{difference imaging}, and has been shown to be effective, even in fields that are crowded or associated with highly non-uniform unresolved surface brightness \citep{Tomaney1996DiffImaging,Bond2001RealtimeDiffImaging}. Most transient surveys, including ZTF, use this method, and `trigger' a transient event when there is a detection in a difference image that exceeds a $5\sigma$ signal-to-noise (S/N) threshold. Throughout this work, we use \textit{trigger} to identify this time of detection. We refer to early classification as classification made within $2$ days of this trigger, and full classification as classifications made after $40$ days since trigger.

        \subsection{Selection Criteria}
        \label{sec:selection_criteria}
        To create a good and clean training sample, we made a number of cuts before processing the light curves. The selection criteria is described as follows.
        
        \begin{description}
            \item[$z < 0.5$ and $z \neq 0$] \hfill \\
                Firstly, we cut objects with host-galaxy redshifts $z = 0$ or $z > 0.5$ such that all galactic objects and any higher redshift objects were removed as these candidates are too faint to be useful for the early-time progenitor studies that motivated the development of this classifier in the first place. While our work relies on knowing the redshift of each transient, in this low redshift range, we should be able to obtain a redshift from the host galaxy from existing catalogs.
            \item[Sufficient data in the early light curve]
                Next, we ensured that the selected light curves each had at least three measurements before trigger, and at least two of these were in different passbands. Even if these measurements were themselves insufficient to cause a trigger, they help establish a baseline flux. This cut therefore removes objects that triggered immediately after the beginning of the observing season, as these are likely to be unacceptably windowed.
            \item[$b > 15^{\circ}$ and $b < -15^{\circ}$] \hfill \\
                Any object in the galactic plane, with latitude $-15^{\circ} < b < 15^{\circ}$, was also cut from the dataset because our analysis only considers extragalactic transients.
            \item[Selected only transient objects] \hfill \\
            	Finally, while the PLAsTiCC simulations included a range of variable objects, including AGN, RR Lyrae, M-dwarf flares, and Eclipsing Binary events, we removed these from the simulated dataset. This cut on the dataset was made because these long-lived variable candidates will likely be identified to a very high completeness over the redshift range under consideration, and will not be misidentified as a class of astrophysical interest for early-time studies.
        \end{description}

    \subsection{Preprocessing}
    \label{sec:Preprocessing}
        Arguably, one of the most important aspects in an effective learning algorithm is the quality of the training set. In this section we discuss efforts to ensure that the data is processed in a uniform and systematic way before we train our DNN.
        
        The light curves are measured in flux units, as is expected for the ZTF difference imaging pipeline. The simulations have a significant fraction of the observations being 5-10 sigma outliers. These outliers are intended to replicate the difference image analysis artifacts, telescope CCD deficiencies, and cosmic rays seen in observational data. We perform `sigma clipping' to reject these outliers. We do this by rejecting photometric points with flux uncertainties that are more than $3\sigma$ from the mean uncertainty in each passband, and iteratively repeat this clipping $5$ times. Next, we correct the light curves for interstellar extinction using the reddening function of \citet{Fitzpatrick1998}. We assume an extinction law, $R_V=3.1$, and use the central wavelength of each ZTF filter to de-redden each light curve listed as follows\footnote{We use the \texttt{extinction} code: \url{https://extinction.readthedocs.io}}:
        \begin{center}
        $g$: 4767 \AA,  $r$: 6215 \AA.
        \end{center}
        
        Following this, we account for cosmological time dilation using the host redshifts, $z$, and convert the observer frame time since trigger to a rest-frame time interval,
        \begin{equation}
            t = (T_{\text{obs}} - T_{\text{trigger}}) / (1+z),
        \end{equation}
        where capital $T$ refers to an observer frame time in MJD and lowercase $t$ refers to a rest-frame time interval relative to trigger. We define trigger as the epoch at which the ZTF difference imaging detects a $5\sigma$ threshold change in flux.
        
        We then calculate the luminosity distance, $d_L(z)$, to each transient using the known host redshift and assuming a $\Lambda$CDM cosmology with $\Omega_M = 0.3$, $\Omega_\Lambda = 0.7$ and $H_0 = 70$. We correct the flux for this distance by multiplying each flux by $d_L^2$ and scaling by some normalizing factor, $\text{norm}=10^{18}$, to keep the flux values in a good range for floating-point machine precision. A measure for the distance-corrected flux, which is proportional to luminosity, is
        \begin{equation}
            L_{\mathrm{data}} (t) = \left( F(t) - F(t)_{\mathrm{med}} \right) \cdot \frac{{d^2_L}(z)}{\mathrm{norm}},
        \end{equation}
        where $F(t)$ is the raw flux value and $F(t)_{\text{med}}$ is the median value of the raw flux points that were observed before the trigger. This median value is representative of the background flux. 
        Even for objects observed by a single survey, with a common set of passbands on a common photometric system, comparing the fluxes of different sources in the same rest-frame wavelength range requires that the light curve photometry be transformed into a common reference frame, accounting for the redshifting of the sources. However, this $k$-correction~\citep{kcorr} requires knowledge of the underlying spectral energy distribution (SED) of each source, and therefore its type --- the goal of this work. 
        Therefore, we have not $k$-corrected these data into the rest-frame, and hence, $L_\mathrm{data}$ cannot be considered the true rest-frame luminosity in each passband.
        
        \subsubsection{Modeling the Early Light Curve}
        \label{sec:tsquare}
        
            \begin{figure}
        	\centering
        	\includegraphics[width=1.\linewidth]{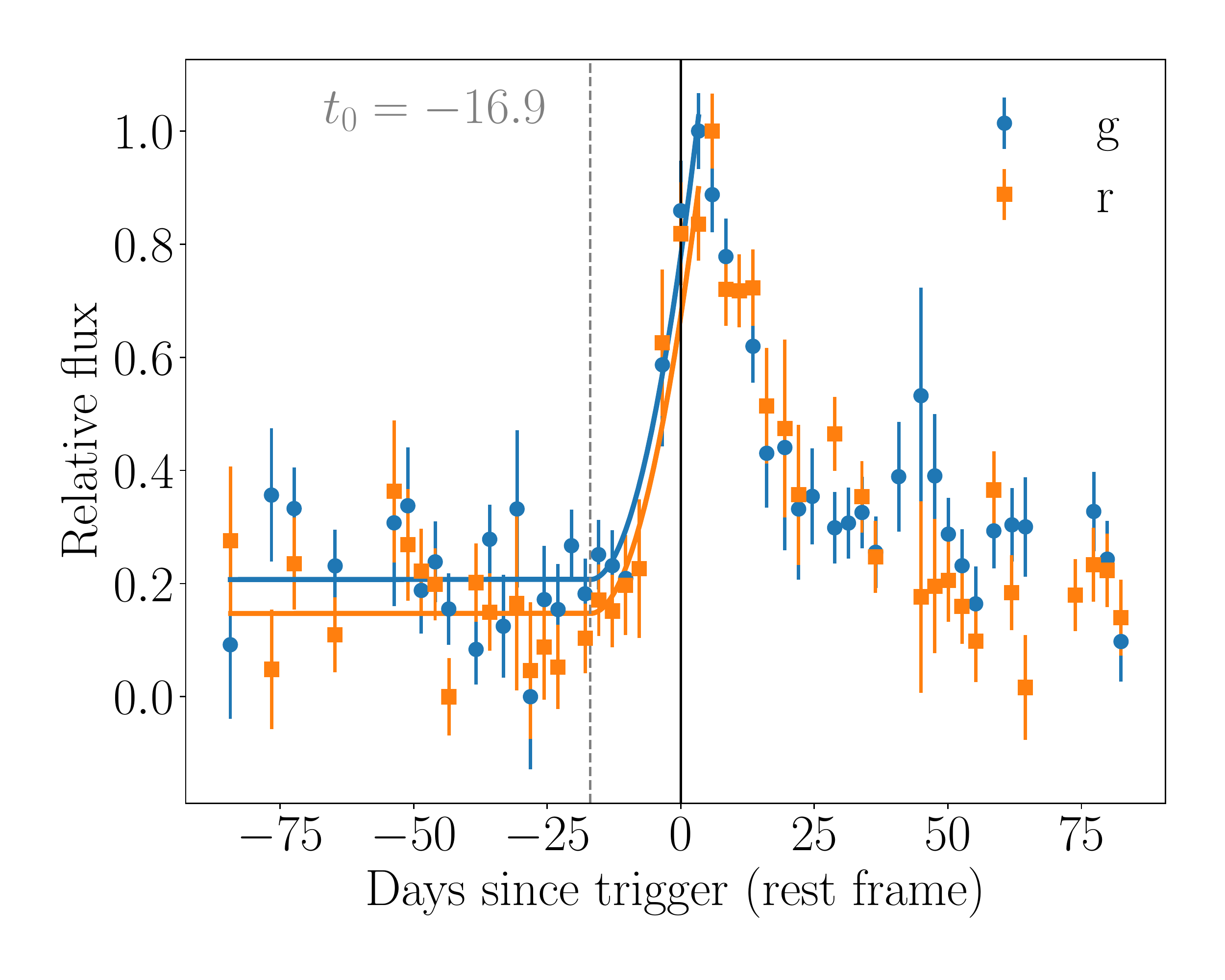}
        	\caption{An example preprocessed Type Ia Supernova light curve from the ZTF simulated dataset (redshift$=0.174$). The normalized fluxes of the $r$ and $g$ passbands are plotted with errors and the solid line is the best fit model of the pre-maximum part the light curve (up to $t_{\text{peak}}$) using equations \ref{eq:model_early} and \ref{eq:chi2_model_early}. The horizontal axis is plotted in the rest-frame (redshift corrected), while the vertical axis is the relative de-reddened and distance-corrected flux (or relative luminosity). The vertical black solid line is the date that difference imaging records a trigger alert, and the vertical grey dashed line is the model's prediction of the explosion date with respect to trigger.}
        	\label{fig:early_lightcurve}
        	\end{figure}
        	
            The ability to predict the class of an object as a function of time is one of the main advantages of \texttt{RAPID} over previous work. Critical to achieving this is determining the epoch at which transient behaviour begins (usually the date of explosion) so that we can \textit{teach} our DNN what a pre-explosion looks like. Basic geometry suggests that a single explosive event should evolve in flux proportional to the square of time \citep{Arnett1982TypeCurve}. While future work might try to fit a power law, we are limited by the sparse and noisy data in the early light curve. Therefore, we model the pre-maximum part of each light curve in each passband, $\lambda$, with a simple $t^2$ fit as follows,
	        \begin{equation}
	            L_\mathrm{mod}^{\lambda} (t; t_0, a^{\lambda}, c^{\lambda}) = \left[a^{\lambda} (t-t_0)^2\right] \cdot H(t-t_0) + c^{\lambda},
	            \label{eq:model_early}
	        \end{equation}
	        where $L_\mathrm{mod} (t)$ is the modelled relative luminosity, $t_0$ is the estimate for time of the initial explosion or collision event, $H(t-t_0)$ is the Heaviside function, and $a$ and $c$ are constants for the amplitude and intercept of the early light curve. The Heaviside function is used to model the $t^2$ relationship after the explosion, $t_0$, and fit a constant flux, $c$, before $t_0$. We define the pre-maximum part of the light curve as observations occurring up to the simulated peak luminosity of the light curve, $t_{\text{peak}}$. We emphasize that this early model is only required for the training set, and therefore, we are able to use the simulated peak time, which will not be available on observed data and is not used for the testing set.
	        
	        We make the assumption that the light curves from each passband have the same explosion date, $t_0$, and fit the light curves from all passbands simultaneously. This is a 5 parameter model: two free parameters, slope ($a^{\lambda}$) and intercept ($c^{\lambda}$), for each of the two passbands, and a shared $t_0$ parameter. We aim to optimize the model's fit to the light curve by first defining the chi-squared for each transient as:
	        \begin{equation}
	            \chi^2 (t_0, \mathbf{a}, \mathbf{c}) = \sum_{\lambda} \sum_{t=-\infty}^{t_{\text{peak}}} \frac{[L_\mathrm{data}^{\lambda}(t) - L_\mathrm{mod}^{\lambda}(t; t_0, a^{\lambda}, c^{\lambda})]^2}{\sigma^{\lambda}(t)^2},
	            \label{eq:chi2_model_early}
	        \end{equation}
	        where $\lambda$ is the index over passbands, $\sigma(t)$ are the photometric errors in $L_\mathrm{data}$, and the sum is taken over all observations at the position of the transient until the time of the peak of the light curve, $t_{\text{peak}}$.
	
	        We sampled the posterior probability $\propto \exp{(-\frac{1}{2} \chi^2)}$ using MCMC (Markov Chain Monte Carlo) with the affine-invariant ensemble sampler as implemented in the Python package, \texttt{emcee} \citep{Foreman-Mackey2013EmceeHammer}. We set a flat uniform prior on $t_0$ to be in a reasonable range before trigger, $-35 < t_0 < 0$, and have a flat improper prior on the other parameters. We use 200 walkers and set the initial positions of each walker as a Gaussian random number with the following mean values: the median of $L_\mathrm{data}^{\lambda}$ for $c^{\lambda}$, the mean of the $L_\mathrm{data}^{\lambda}$ for $a^{\lambda}$, and $-12$ for $t_0$. We ran each walker for $700$ steps, which after analyzing the convergence of a few MCMC chains, we deemed reasonable to not be too computationally expensive while still finding approximate best fits for $\mathbf{a}, \mathbf{c}$ and $t_0$. The best fit early light curve for an example Type Ia supernova in the training set is illustrated in Fig.~\ref{fig:early_lightcurve}.

    We summarize the selection criteria and preprocessing stages applied to the testing and training sets as detailed in sections \ref{sec:selection_criteria} - \ref{sec:Preprocessing} in Table \ref{tab:summaryCuts}.
    
    \begin{table}[]
        \centering
        \begin{tabular}{c|c}
            \hline
             Selection Criteria & Applied To \\ \hline 
             $0 < z \leq 0.5$ & Train \& Test \\ 
             $b < 15^{\circ}$ or $b > 15^{\circ}$  & Train \& Test \\ 
             At least 3 points pre-trigger  & Train \& Test \\ 
             \hline \hline
             Preprocessing & Applied to  \\ \hline 
             Sigma-clipping fluxes & Train \& Test \\ 
             Undilate the light-curves by $(1+z)$ & Train \& Test \\ 
             Correct light curves for distance. & Train \& Test \\
             Correct Milky Way extinction & Train \& Test \\ 
             Rescale light curves between 0 and 1 & Train \& Test \\ 
             Model early light curve to obtain $t_0$ & Train only\footnote{Applied to training set for designating pre-explosion. Applied to test set to evaluate performance.} \\ 
             Keep only $-70 < t < 80$ days from trigger & Train \& Test \\ 
             \hline
        \end{tabular}
        \caption{Summary of the cuts and preprocessing steps applied to the training and testing sets. The selection criteria help match the simulations to what we expect from the observed ZTF data-stream.}
        \label{tab:summaryCuts}
    \end{table}
    
    \subsection{Training Set Preparation}
        \label{sec:FinalDataPrep}
        Irregularly sampled time-series data is a common problem in machine learning, and is particularly prevalent in astronomical surveys where the intranight cadence choices and seasonal constraints lead to naturally arising temporal gaps. Therefore, once the light curves have been processed and $t_0$ has been computed for each transient, we linearly interpolate between the unevenly sampled time series data. From this interpolation, we impute data points such that each light curve is sampled at 3-day intervals between $-70 < t < 80$ days since trigger (or as far as the observations exist), to give a vector of length $n=50$, where we set the values outside the data range to zero. We ensure that each light curve in a given passband is sampled on the same 3-day grid. The final input image for each transient $s$ is $\bm{I}^s$, which is a matrix with each row composed of the imputed light curve fluxes for each passband and two additional rows containing repeated values of the host-galaxy redshift in one row and the MW dust reddening in the other row. Hence, the input image is an $n \times (p+2)$ matrix, where $p$ is the number of passbands. This input image, $\bm{I}^s$, is illustrated as the \textit{Input Matrix} in Fig.~\ref{fig:matrix_schematic}.
        
        One of the key differences in this work compared to previous light curve classification approaches is our ability to provide time-varying classifications. Key to computing this, is labelling the data at each epoch rather than providing a single label to an entire light curve. Using the value of $t_0$ computed in section \ref{sec:tsquare}, we define two phases of each transient light curve: the pre-explosion phase (where $t < t_0$), and the transient phase (where $t \geq t_0$). Therefore, the label for each light curve is a vector of length $n$ identifying the transient class at each time-step. This $n$-length vector is subsequently one-hot encoded, such that each class is changed to a zero-filled vector with one element set to $1$ to indicate the transient class (see equation \ref{eq:onehot}). This transforms the $n$-length label vector into an $n \times (m+1)$ vector, where $m$ is the number of transient classes. This is illustrated as the \textit{Class Matrix} in Fig.~\ref{fig:matrix_schematic}.

\section{Model}
    \label{sec:Model}
    \subsection{Framing the Problem}
        \label{sec:Framing}
        In this work, we train a deep neural network (DNN) to map the light curve data of an individual transient $s$ onto probabilities over classes $\{c=1, \ldots, (m+1)\}$.  The DNN models a function that maps an input multi-passband light curve matrix, $\bm{I}^{st}$, for transient $s$ up to a discrete time $t$, onto an output probability vector,
        \begin{equation}
            \bm{y}^{st} = \bm{f}_t(\bm{I}^{st}; \bm{\theta}), 
        \end{equation}
        where $\bm{\theta}$ are the parameters (e.g. weights and biases of the neurons) of our DNN architecture. We define the input $\bm{I}^{st}$ as an $n \times (p+2)$ matrix\footnote{The reader can consider $\bm{I}^{st}$ as an image that zeros out all future fluxes after a time $t$, hence preserving the $n \times (p+2)$ matrix shape irrespective of the image phase coverage. The function $\bm{f}_t(\cdot\,; \bm{\theta})$ only uses the information in the input light curve up to time $t$.} representing the light curve up to a time-step, $t$. The output $\bm{y}^{st}$ is a probability vector with length $(m+1)$, where each element $y^{st}_c$ is the model's predicted probability of each class $c$ (at each time step), such that $y^{st}_c \ge 0$ and $\sum^{m+1}_{c=1} y^{st}_c = 1.$ 
        
        First, to quantify the discrepancy between the model probabilities and the class labels we define a weighted \textit{categorical cross-entropy},
        \begin{equation}
        H_{w} (\bm{Y}^{st}, \bm{y}^{st}) = -\sum\limits_{c=1}^{m+1} w_c \, Y^{st}_c \log(y^{st}_c),
        \label{eq:cross-entropy}
        \end{equation}
        where $w_c$ is the weight of each class, $\bm{Y}^{st}$ is the label for the true transient class at each time-step and is a one-hot encoded vector of length $(m+1)$ such that,
        \begin{equation}
        \label{eq:onehot}
         Y^{st}_c =
            \begin{cases}
                1 & \text{if $c$ is the true class of transient $s$ at time $t$} \\
                0 & \text{otherwise}
            \end{cases}
        \end{equation}
        where the label, $\bm{Y}^{st}$, has two phases, the pre-explosion phase with class $c=1$ when $t < t_0$ and the transient phase with class $c>1$ when $t \geq t_0$.
        
        If weights were equal for all classes, Eq. \ref{eq:cross-entropy} is proportional to the negative log-likelihood of the probabilities of a categorical distribution (or a generalized Bernoulli distribution). However, to counteract imbalances in the distribution of classes in the dataset which may cause more abundant classes to dominate in the optimization, we define the weight for each class $c$ as
        \begin{equation}
            w_c = \frac{N \times n}{N_c},
            \label{eq:weights}
        \end{equation}
        where $N_c$ is the number of times a particular class appears in the $N \times n$ training set.
        
         We define the global objective function as 
        \begin{equation}
            \mathrm{obj}(\bm{\theta}) = \sum\limits_{s=1}^{N}\sum\limits_{t=0}^{n} H_{w} (\bm{Y}^{st}, \bm{y}^{st}),
            \label{eq:objective}
        \end{equation}
        where we sum the weighted categorical cross-entropy over all $n$ time-steps and $N$ transients in the training set. To train the DNN and determine optimal values of its parameters $\bm{\hat{\theta}}$, we minimize this objective function with the sophisticated and commonly used \texttt{Adam} gradient descent optimiser \citep{Kingma2014}. The model $\bm{f}_t(\bm{I}^{st}; \bm{\hat{\theta}})$ is represented by the complex DNN architecture illustrated in Fig.~\ref{fig:RNN_architecture} and is described in the following section.
                
\subsection{Recurrent Neural Network Architecture}
    \label{sec:RNN}
    
    \begin{figure*}[htbp]
	\centering
	\includegraphics[width=1.0\linewidth]{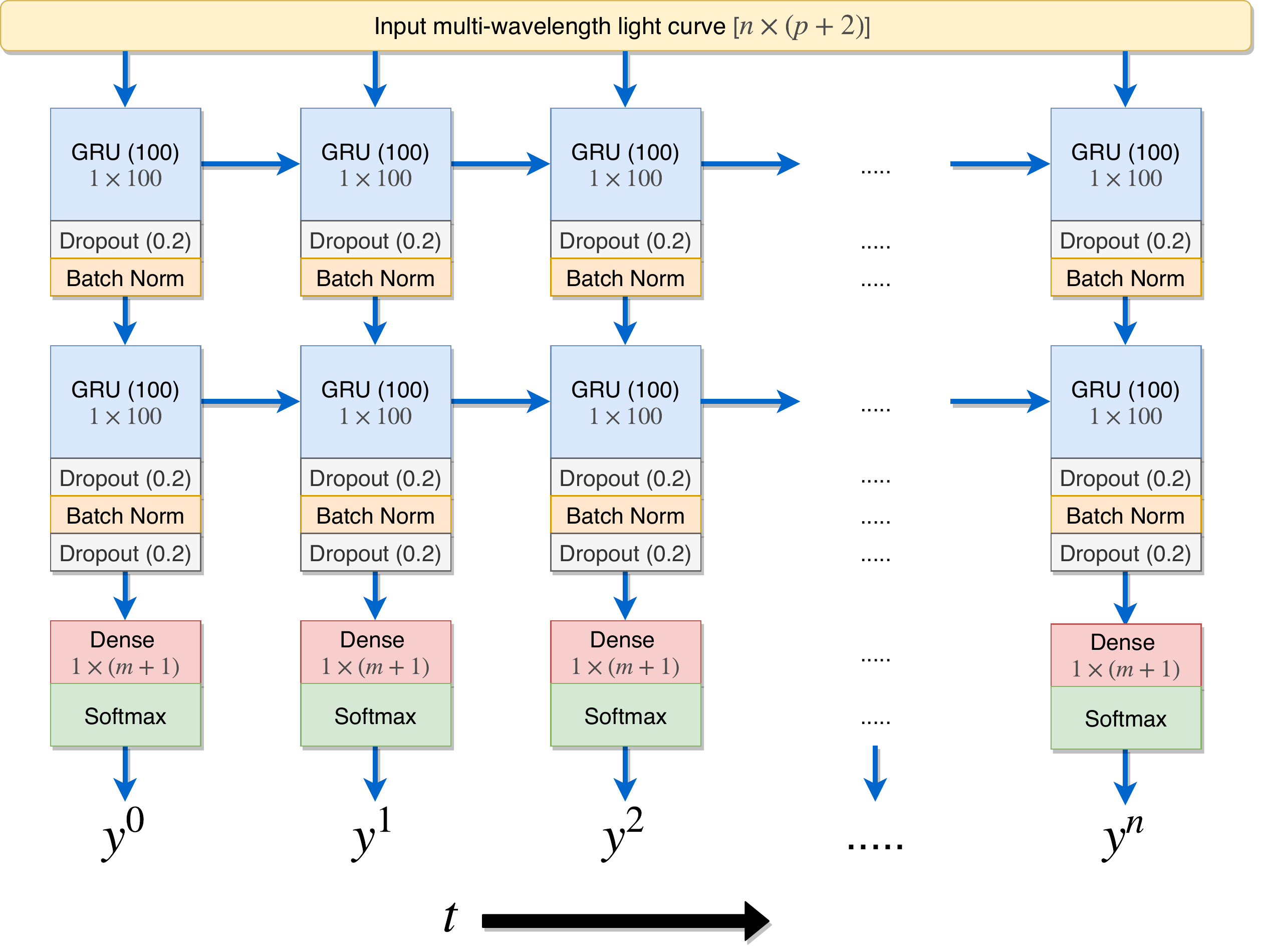}
	\caption{Schematic of the deep recurrent neural network architecture used in {\tt RAPID}. Each column in the diagram is one of the $n$ time steps of the processed light curve, while each row represents a different neural network layer. The grey text in each block states the shape of the output matrix of each layer in that block. The input image is composed of an $n\times (p+2)$ matrix consisting of the light curve fluxes, host redshift, and Milky Way reddening. Two uni-directional gated recurrent unit layers of size $100$ are used for encoding and decoding the input sequences, respectively. It is in these RNN layers that information about previous time-steps is encoded. Batch normalization is applied between each layer to normalize the network parameters and hence, speed the training process.  To counter overfitting during training, we employ the dropout optimization technique \citep{Srivastava2014} to the neurons in each of the GRU and Batch Normalization layers, and set the dropout rate to 20\%. Finally, a fully-connected (dense) layer with a softmax regression activation function is applied to compute the probability of each class at each time-step. We wrap the final layer in \texttt{Keras}' \textit{Time Distributed} layer so that each time step is treated independently, and only uses information from the current and previous time-steps.}
	\label{fig:RNN_architecture}
	\end{figure*}  
    
    Recurrent Neural Networks (RNNs), such as Long Short-Term Memory (LSTM) and Gated Recurrent Unit (GRU) networks have been shown to achieve state-of-the-art performance in many benchmark time-series and sequential data applications \citep{Bahdanau2014NeuralTranslate,Sutskever2014SequenceNetworks,Che2018RecurrentValues}. Its success in these applications is due to its ability to retain an internal memory of previous data, and hence capture long-term temporal dependencies of variable-length observations in sequential data. \textbf{We extend this architecture to our case with a time-varying multi-channel (multiple passbands) input and a time-varying multi-class output.}
    
    Recently, \citet{Naul2018AStars}, \citet{Charnock2016}, \citet{Moss2018}, and \citet{Hinners2017MachineClassification} have used two RNN layers for this framework on astronomical time-series data. However, our work differs from these by making use of \textit{uni-directional} GRUs instead of \textit{bi-directional} RNNs. Bi-directional RNNs are able to pass information both forwards and backwards through the neural network representation of the light curve, and can hence preserve information on both the past and future at any time-step. However, this is only suitable for retrospective classification, because it requires that we wait for the entire light curve to complete before obtaining a classification. The real-time classification used in our work is a novel approach in time-domain astronomy, but necessitates the use of uni-directional RNNs. Hence, our two RNN layers read the light curve chronologically. 

    The deep neural network (DNN) is illustrated in Fig.~\ref{fig:RNN_architecture}. We have developed the network with the high level \texttt{Python} API, \texttt{Keras} \citep{Keras}, built on the recent highly efficient \texttt{TensorFlow} machine learning system \citep{Abadi2015}. We describe the architecture in detail here.
    \begin{description}
        \item[Input] As detailed in section \ref{sec:FinalDataPrep}, the input is an $n \times (p+2)$ matrix. However, as we are implementing a sequence classifier, we can consider the input at each time-step as being vector of length $(p+2)$.
        
        \item[First GRU Layer] Gated Recurrent Units are an improved version of a standard RNN and are a variation of the LSTM (see \citealt{GRUvsLSTM} for a detailed comparison and explanation). We have selected GRUs instead of LSTMs in this work, as they provide appreciably shorter overall training time, without any significant difference in classification performance. Both are able to capture long-term dependencies in time-varying data with parameters that control the information that should be remembered at each step along the light curve. We use the first GRU layer to read the input sequence one time-step at a time and encode it into a higher-dimensional representation. We set-up this GRU layer with 100 units such that the output is a vector of shape $1 \times 100$.
        
        \item[Second GRU Layer] The second GRU layer is conditioned on the input sequence. It takes the output of the previous GRU and generates an output sequence. Again, we use 100 units in the GRU to maintain the $n \times 100$ output shape. We use uni-directional GRUs that enable only information from previous time-steps to be encoded and passed onto future time-steps.
        
        \item[Batch Normalization]
            We then apply \textit{Batch Normalization} (first introduced in \citealt{BatchNorm}) to each GRU layer. This acts to improve and speed up the optimization while adding stability to the neural network and reducing overfitting. While training the DNN, the distribution of each layer's inputs changes as the parameters of the previous layers change. To allow the parameter changes during training to be more stable, batch normalization scales the input. It does this by subtracting the mean of the inputs and then dividing it by the standard deviation.
        \item[Dropout] We also implement dropout regularization to each layer of the neural network to reduce overfitting during training. This is an important step that effectively ignores randomly selected neurons during training such that their contribution to the network is temporarily removed. This process causes other neurons to more robustly handle the representation required to make predictions for the missing neurons, making the network less sensitive to the specific weights of any individual neuron. We set the dropout rate to 20\% of the neurons present in the previous layer each time the Dropout block appears in the DNN in Fig.~\ref{fig:RNN_architecture}.
        
        \item[Dense Layer] A dense (or fully-connected) layer is the simplest type of neural network layer. It connects all 100 neurons at each time-step in the previous layer, to the $(m+1)$ neurons in the output layer simply using equation \ref{eq:neuron}. As this is a classification task, the output is a vector consisting of all $m$ transient classes and the \textit{Pre-explosion} class. However, as we are interested in time-varying classifications, we wrap this Dense layer with a \textit{Time-Distributed} layer, such that the dense layer is applied independently at each time-step, hence giving an output matrix of shape $n \times (m+1)$.
        
        \item[Neurons]
            The output of each neuron in a neural network layer can be expressed as the weighted sum of the connections to it from the previous layer:
                \begin{equation}
                    \hat{y}_i = f\left(\sum\limits_{j=1}^{M} W_{ij} \, x_j + b_i \right),
                    \label{eq:neuron}
                \end{equation}
            where $x_j$ are the different inputs to each neuron from the previous layer, $W_{ij}$ are the weights of the corresponding inputs, $b_i$ is a bias that is added to shift the threshold of where inputs become significant, $j$ is an integer running from 1 to the number of connected neurons in the previous layer ($M$), and $i$ is an integer running from 1 to the number of neurons in the next layer. For the Dense layer, $\mathbf{x}$ is simply the $(1\times100)$ matrix from the output of the GRU and Batch Normalisation, $\mathbf{y}$ is made up of the $(m+1)$ output classes, $j$ runs from 1 to ($m+1$) and $i$ runs across the $100$ input neurons from the GRU. The matrix of weights and biases in the Dense layer and throughout the GRU layers are some of the free parameters that are computed by {\tt TensorFlow} during the training process.
        \item[Activation function] 
            As with any neural network, each neuron applies an activation function $f(\cdot)$ to bring non-linearity to the network and hence help it to adapt to a variety of data. For feed-forward networks it is common to make use of Rectified Linear Units \citep[ReLU,][]{Nair2010} to activate neurons. However, the GRU architecture uses sigmoid activation functions as it outputs a value between $0$ and $1$ and can either let no flow or complete flow of information from previous time-steps. 
        \item[Softmax regression] The final layer applies the softmax regression activation function, which generalises the sigmoid logistic regression to the case where it can handle multiple classes. It applies this to the Dense layer output at each time-step, so that the output vector is normalized to a value between 0 and 1 where the sum of the values of all classes at each time-step sums to 1. This enables the output to be viewed as a relative probability of an input transient being a particular class at each time-step. The output probability vector,
            \begin{equation}
            \bm{y} = \mathrm{softmax}(\bm{\hat{y}}),
            \end{equation}
            is computed with a softmax activation function that is defined as
            \begin{equation}
            \mathrm{softmax}(\bm{x})_i = \frac{e^{x_i}}{\sum\limits _j e^{x_j}}.
            \end{equation}
            We use the output softmax probabilities to rank the best matching transient classes for each transient light curve at each time-step.
    \end{description}
    
    We reiterate that the overall architecture is simply a function that maps an input $n\times(p+2)$ light curve matrix onto an $n \times (m+1)$ softmax probability matrix indicating the probability of each transient class at each time-step. In order to optimize the parameters of this mapping function, we specify a weighted categorical cross-entropy loss-function that indicates how accurately a model with given parameters matches the true class for each input light curve (as defined in equation \ref{eq:cross-entropy}). 

    We minimize the objective function defined in equation \ref{eq:objective} using the commonly used, but sophisticated stochastic gradient descent optimizer called the \texttt{Adam} optimizer \citep{Kingma2014}. As the class distribution is inevitably uneven, and the pre-explosion class is naturally over-represented as it appears in each light curve label, we prevent bias towards over-represented classes by applying class-dependent weights while training as defined in equation \ref{eq:weights}.

    The several layers in the DNN create a model that has over one hundred thousand free parameters. As we feed in our training set in batches of 64 light curves at a time, the neural network updates and optimizes these parameters. While the size of the parameter space seems insurmountable, the \texttt{Adam} optimizer is able to compute individual adaptive learning rates for different parameters from estimates of the mean and variance of the gradients and has been shown to be extraordinarily effective at optimizing high-dimensional deep learning models.

    With the often quoted `black box' nature of machine learning, it is always a worry that the machine learning algorithms are learning traits that are specific to the training set but do not reflect the physical nature of the classes more generally. Ideally, we would like to ensure that the model we build both accurately captures regularities in the training data while simultaneously generalizing well to unseen data. Simplistic models may fail to capture important patterns in the data, while models that are too complex may overfit random noise and capture spurious patterns that do not generalize outside the training set. While we implement regularization layers (dropout) to try to prevent overfitting, we also monitor the performance of the classifier on the training and testing sets during training. In particular, we ensure that we do not run the classifier over so many iterations that the difference between the values of the objective function evaluated on the training set and the testing set become significant.
   
\section{Results}
    \label{sec:Performance}

    In this section we detail the performance of {\tt RAPID} trained on simulated ZTF light curves. The dataset consists of 48029 transients split between 12 different classes, where each class has approximately 4000 transients. We trained our DNN on  60\% of this set and tested its performance on the remaining 40\%. The data was preprocessed using the methods outlined in sections \ref{sec:selection_criteria} - \ref{sec:FinalDataPrep}. Processing this set, and then training the DNN on it, was computationally expensive, taking several hours to train. Once the DNN is trained, however, it is able to classify several thousands of transients within a few seconds.

    \subsection{Hyper-parameter Optimization}
        One of the key criticisms of deep neural networks is that they have many hyper-parameters describing the architecture that need to be set before training. As training our DNN architecture takes several hours, optimizing the hyper-parameter space by testing the performance of a range of setup values is a very slow process that most similar work have not attempted. Despite this challenge, we performed a broad grid-search of three of our DNN hyper-parameters: number of neurons in each GRU layer, and the dropout fraction. After testing 12 different setup parameters, we found that there was only a 2\% variation on the overall accuracy. The hyper-parameters that are shown in Fig.~\ref{fig:RNN_architecture} were the best performing set of parameters.

    \subsection{Accuracy}
    
    \begin{figure}[htpb]
	\centering
	\includegraphics[width=1.\linewidth]{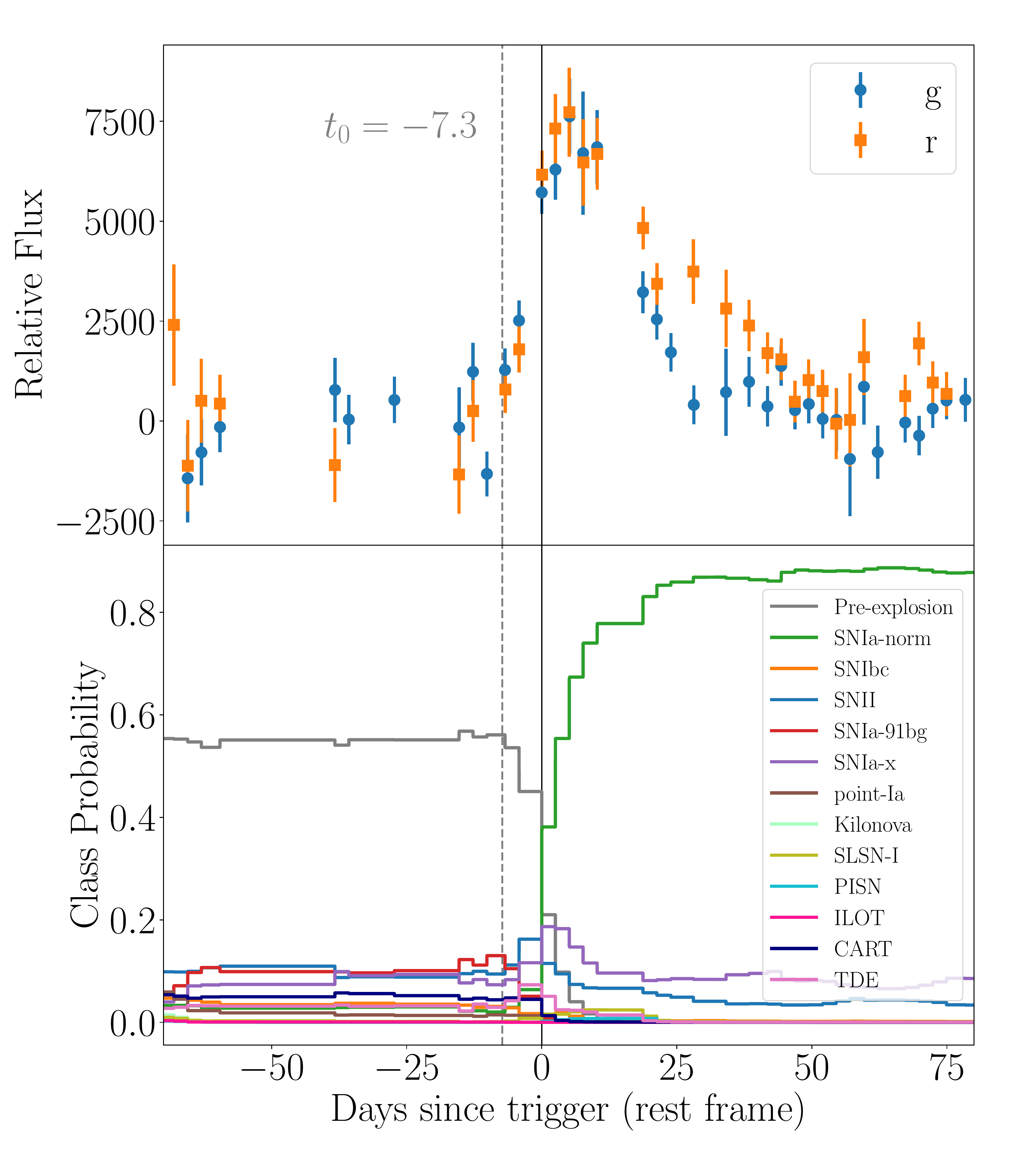}
	\caption{An example normal SNIa light curve from the testing set (redshift$=0.174$) is shown in the top panel, and the softmax classification probabilities from \texttt{RAPID} are plotted as a function of time over the light curve in the bottom panel. The plot shows the rest frame time since trigger. The vertical grey dashed line is the predicted explosion date from our $t^2$ model fit of the early light curve (see section \ref{sec:tsquare}). Initially, the object is correctly predicted to be \textit{Pre-explosion}, before it is more confidently predicted as a SNIa-Normal at -20 days before the trigger. Hence, the neural network predicts the explosion date only 4 days after early light curve model fit's prediction. The confidence in the predicted classification improves over the lifetime of the transient.}
	\label{fig:classification_vs_time}
	\end{figure}
	
    \begin{figure}[htpb]
	\centering
	\includegraphics[width=1.\linewidth]{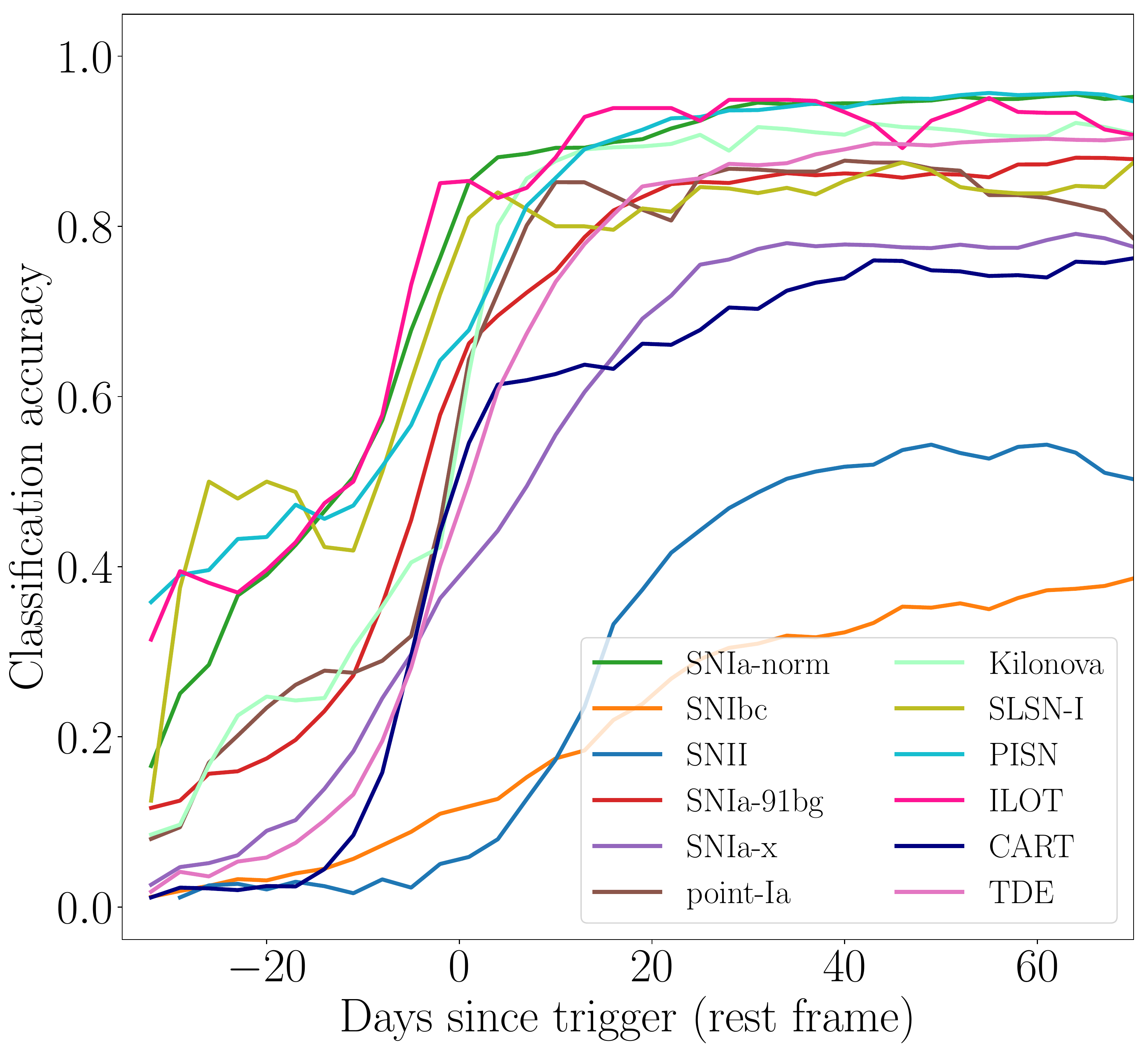}
	\caption{The classification accuracy of each transient class as a function of time since trigger in the rest frame. The values correspond to the diagonals of the confusion matrices at each epoch.}
	\label{fig:accuracy_vs_time}
	\end{figure}
	
    We go beyond previous attempts at photometric classification in two important ways. Firstly, we aim to classify a much larger variety of sparse multi-passband transients, and secondly, and most significantly, we provide classifications as a function of time. An example of this is illustrated in Fig.~\ref{fig:classification_vs_time}. At each epoch along the light curve, the trained DNN outputs a softmax probability of each transient class. As more photometric data is provided along the light curve, the DNN updates the class probabilities of the transient based on the state of the network at previous time-steps plus the new time-series data. Within just a few days of the explosion, and well before the ZTF trigger, the DNN was able to correctly learn that the transient evolved from \textit{Pre-explosion} to a SNIa. 
    
    To assess the performance of {\tt RAPID}, we make use of several metrics. The most obvious metric is simply the accuracy, that is, the ratio of correctly classified transients in each class to the total number of transients in each class. At each epoch along every light curve in the testing set, we select the highest probability class and compare this to the true class. After aligning each light curve by its trigger, we obtained the prediction accuracy of each class as a function of time since trigger. This is plotted in Fig.~\ref{fig:accuracy_vs_time}.

	\begin{figure*}[htpb]
	\gridline{
		\fig{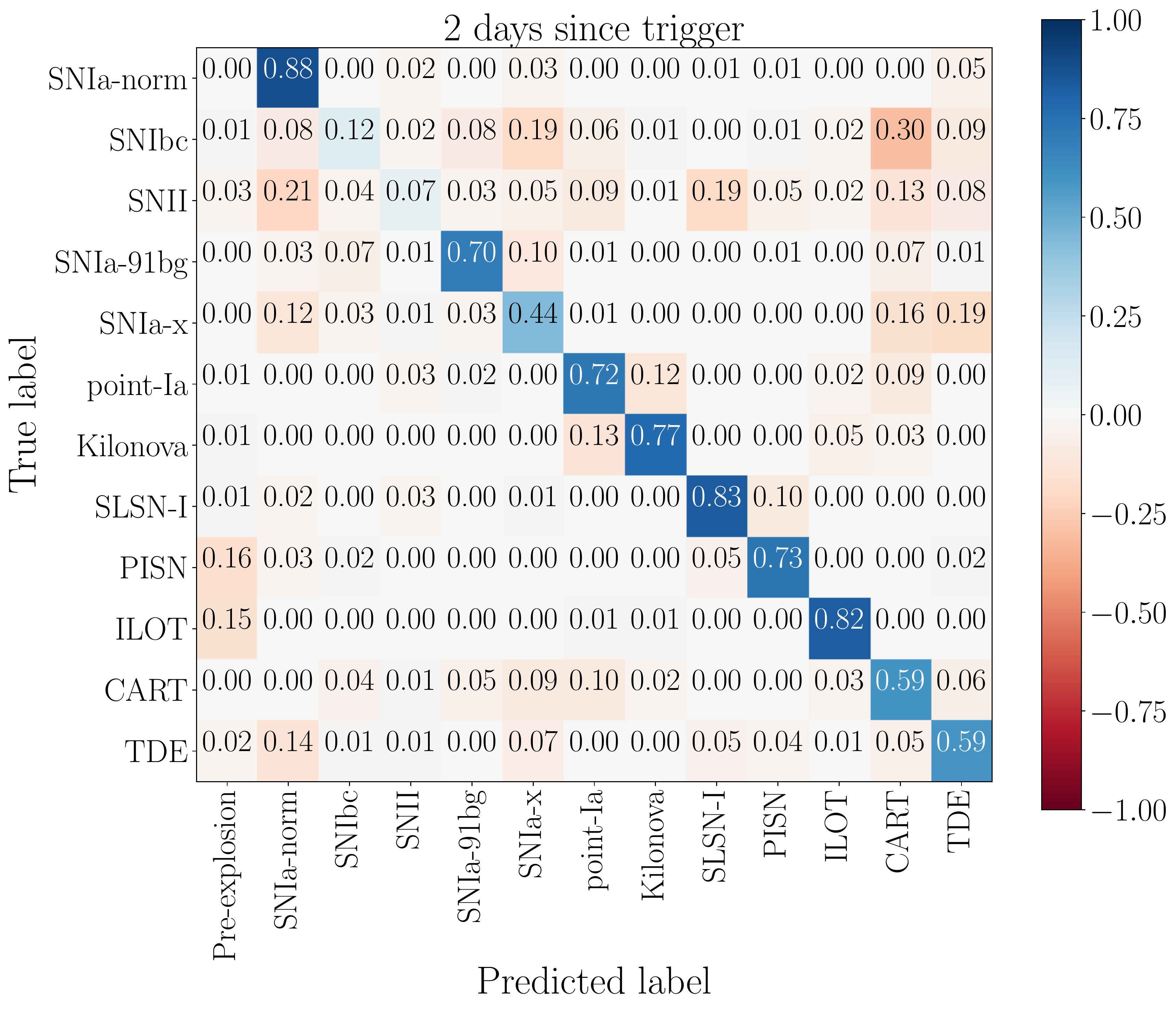}{0.63\textwidth}{(a) Early Epoch}
	}
	\gridline{
		\fig{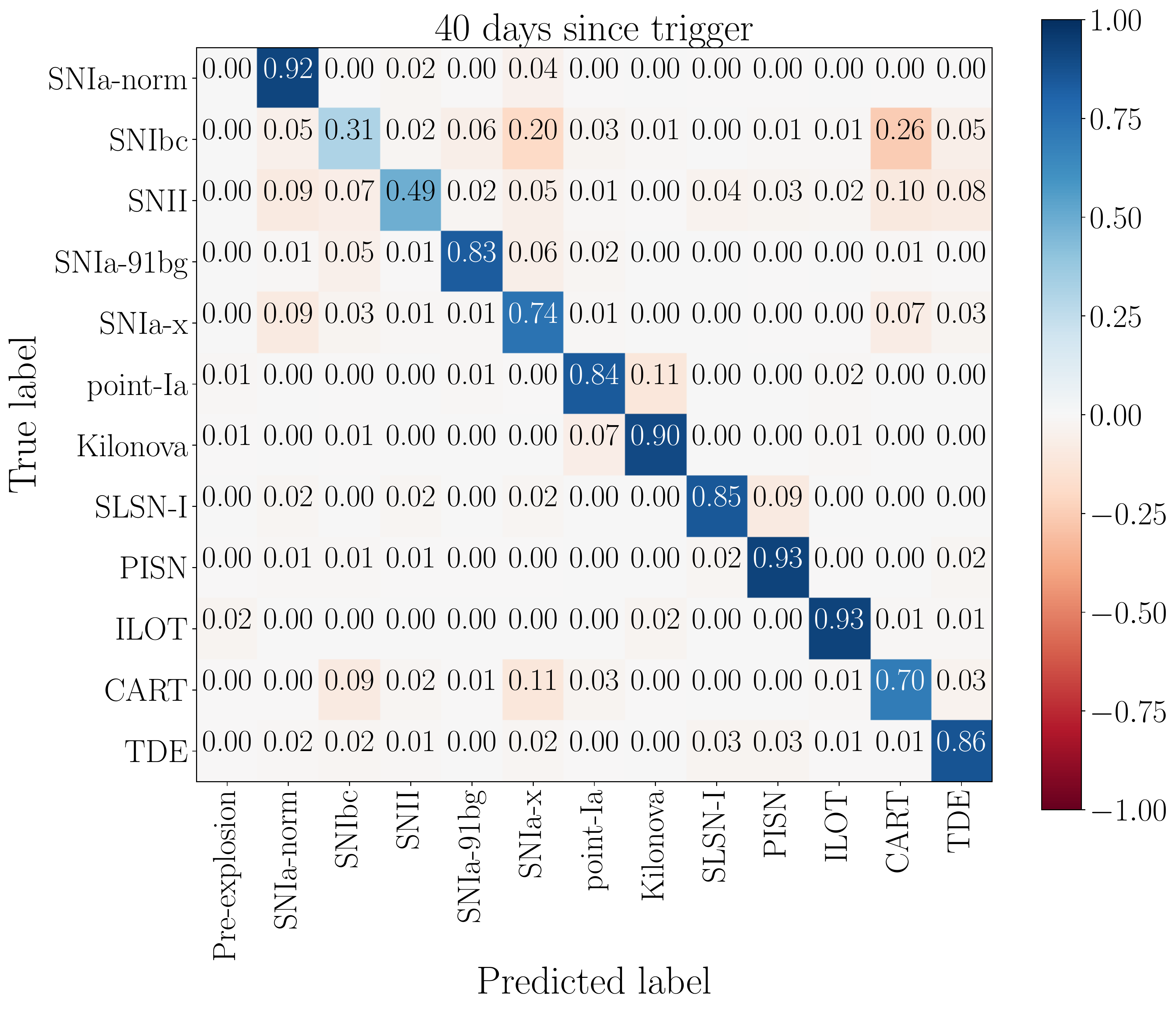}{0.63\textwidth}{(b) Late Epoch}
	}
	\caption{The normalized confusion matrices of the 12 transient classes at 2 days past trigger (top), and at 40 days past trigger (bottom). The confusion matrices show the classification performance tested on 40\% of the dataset after the classifier was trained on 60\% of the dataset. The colour bar and cell values indicate the fraction of each \textit{True Label} that were classified as the \textit{Predicted Label}. Negative colour bar values are used only to indicate misclassifications. Please see the online material (\url{https://www.ast.cam.ac.uk/~djm241/rapid/cf.gif}) for an animation of the evolution of the confusion matrix as a function of time since trigger (showing epochs from -25 to 70 days from trigger).}
	\label{fig:RNN_CF}
    \end{figure*}
    
    The total classification accuracy of each class in the testing set increases quickly before trigger, but then begins to flatten out with only small increases after approximately 20 days post-trigger. For most classes, the transient behaviour of the light curve is generally nearing completion at this stage, and hence we can expect that new photometric data adds little to improving the classification as the brightness tends towards the background flux level. The classification performance of the core-collapse supernovae, SNIbc and SNII, are particularly poor. To better understand this, it is useful to see where misclassifications occurred. 
    
    \subsection{The Confusion Matrix}
    
    The confusion matrix is often a good way to visualize this. Typically, each entry in the matrix describes counts of the number of transients of the true class, $c$, that had the highest predicted probability in class, $\hat{c}$. For ease of interpretability, we make use of a specially normalized confusion matrix in this work. We normalize the confusion matrix such that the ($c$, $\hat{c}$) entry is the fraction of transients of the true class $c$ that are classified into the predicted class $\hat{c}$. With this normalization, each row in the matrix must sum to 1. Therefore each row is an estimate of the classifier's conditional distribution of (maximum probability) predicted labels given each true class label.

    In Fig.~\ref{fig:RNN_CF}, we plot the normalized confusion matrices at an early (2 days post-trigger) and late (40 days post-trigger) stage of the light curve. In the online material, we provide an animation of this confusion matrix evolving in time since trigger (instead of just the two epochs shown here)\footnote{Paper animations can be found here: \url{https://www.ast.cam.ac.uk/~djm241/rapid/}}.
    
\begin{figure}[htpb]
	\centering
	\includegraphics[width=1.\linewidth]{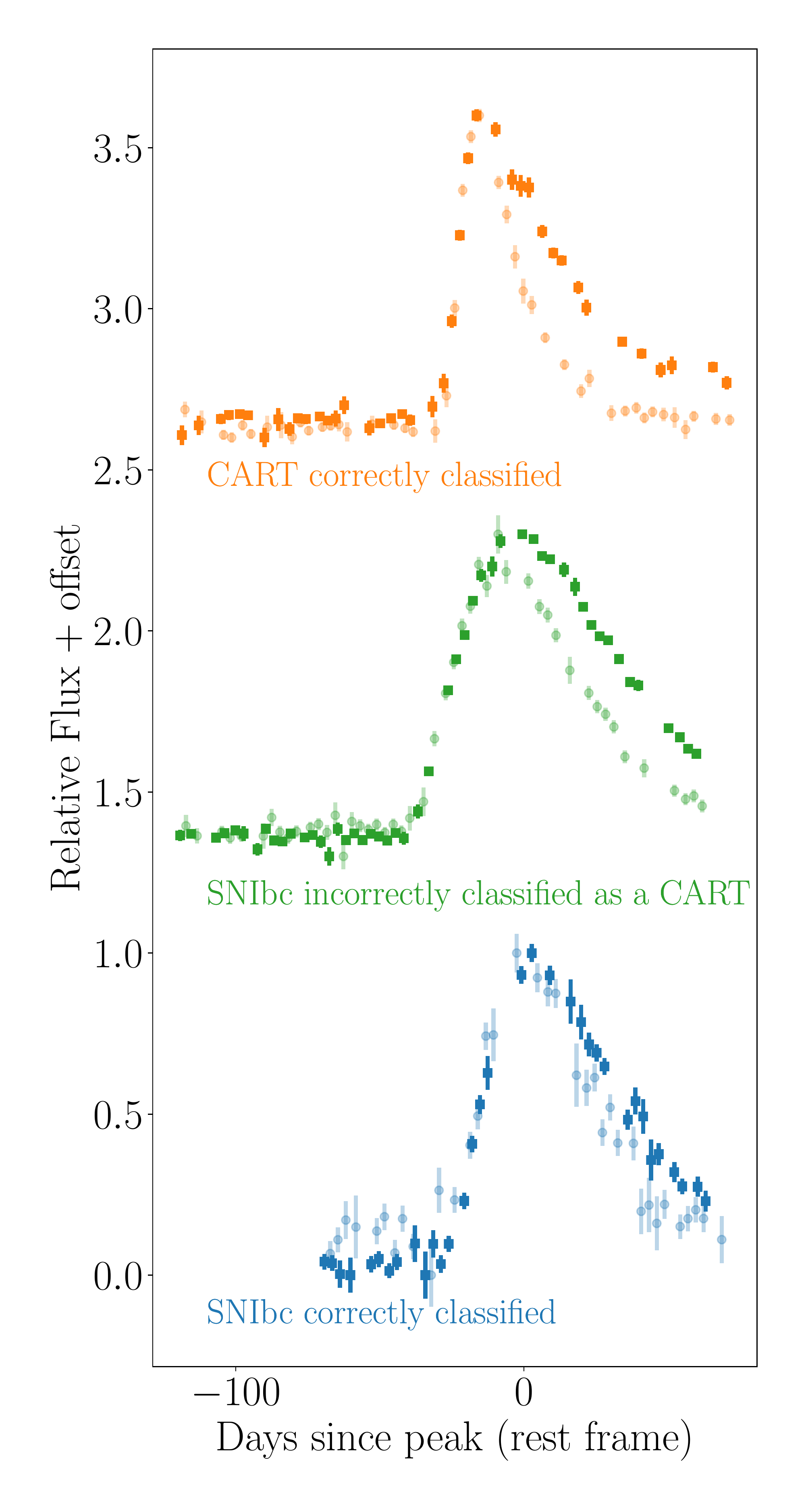}
	\caption{Three of the simulated light curves from our sample - a correctly classified CART (top), a SN Ibc incorrectly classified as a CART (middle), and a correct classified SN Ibc (bottom). The dark-coloured square markers are the $r$ band fluxes and the lighter-coloured circle markers are the $g$ band fluxes. Our classifier is sensitive to light curve shape, and the limited colour information available with ZTF leads to a significant fraction of SN Ibc objects being misclassified as CARTs. We expect classification performance to improve with LSST, which will provide \textit{ugrizy} light curves.}
	\label{fig:ccvscart}
	\end{figure}

    The overall classification performance is, as expected, slightly better at the late phase of the light curve. However, the performance only 2 days after trigger is particularly promising for our ability to identify transients at early times to gather a well-motivated follow-up candidate list. SNe Ia have the highest classification accuracy at early times with most misclassification occurring with other subtypes of Type Ia supernovae. At late times, the Intermediate Luminosity Transients and TDEs are best identified. The core-collapse supernovae (SNIbc, SNII) appear to be most often confused with calcium-rich transients and other supernova types. CARTs are a newly discovered class of transients and their physical mechanism is not yet well-understood. However, the reason for the confusion most likely stems from their fast rise-times similar to many core-collapse SNe. This is illustrated in Fig.~\ref{fig:ccvscart}.  
    
    \subsection{Receiver Operating Characteristic Curves}

\begin{figure*}[htpb]
	\gridline{
		\fig{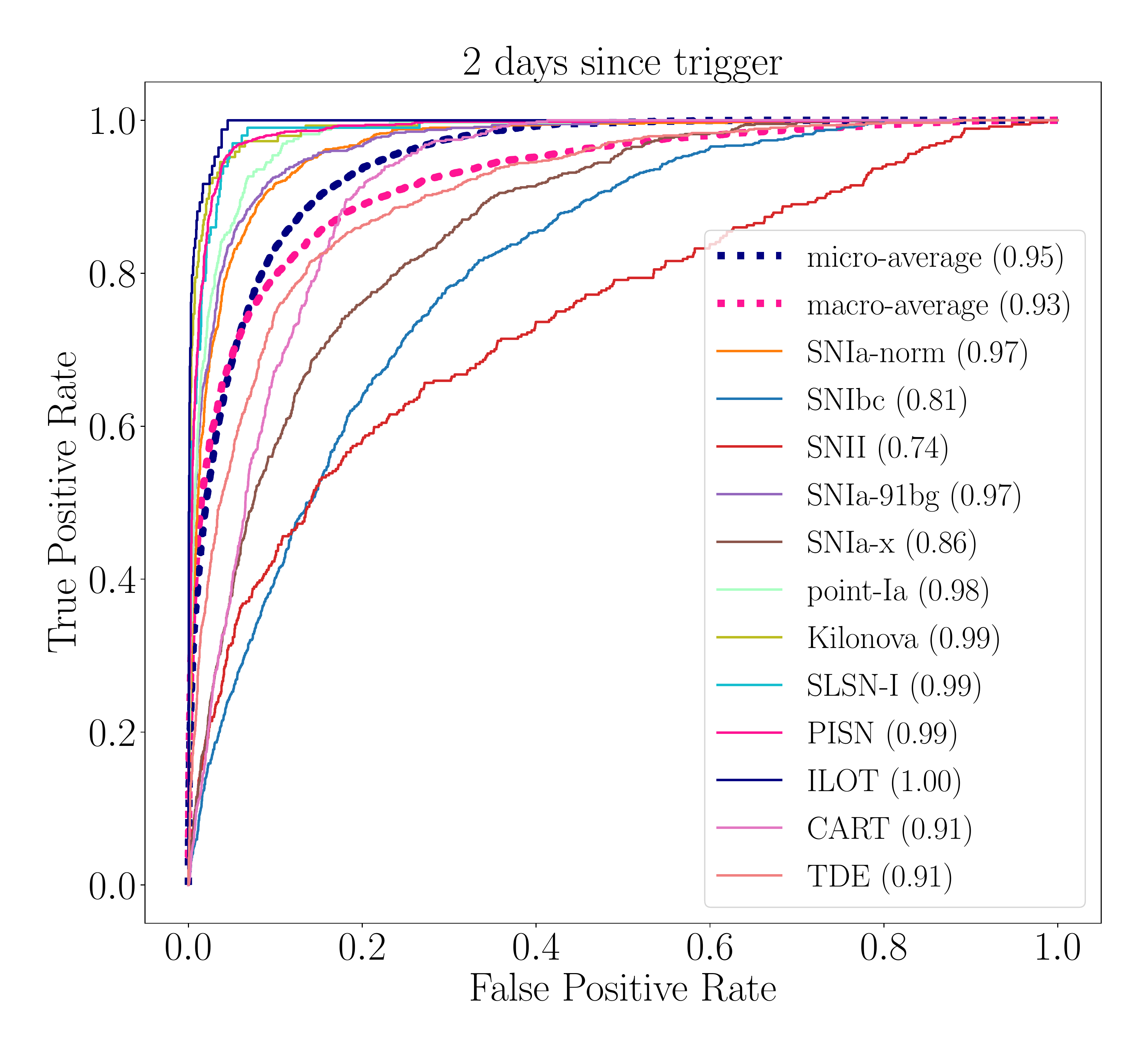}{0.5\textwidth}{(a) Early Epoch}
		\fig{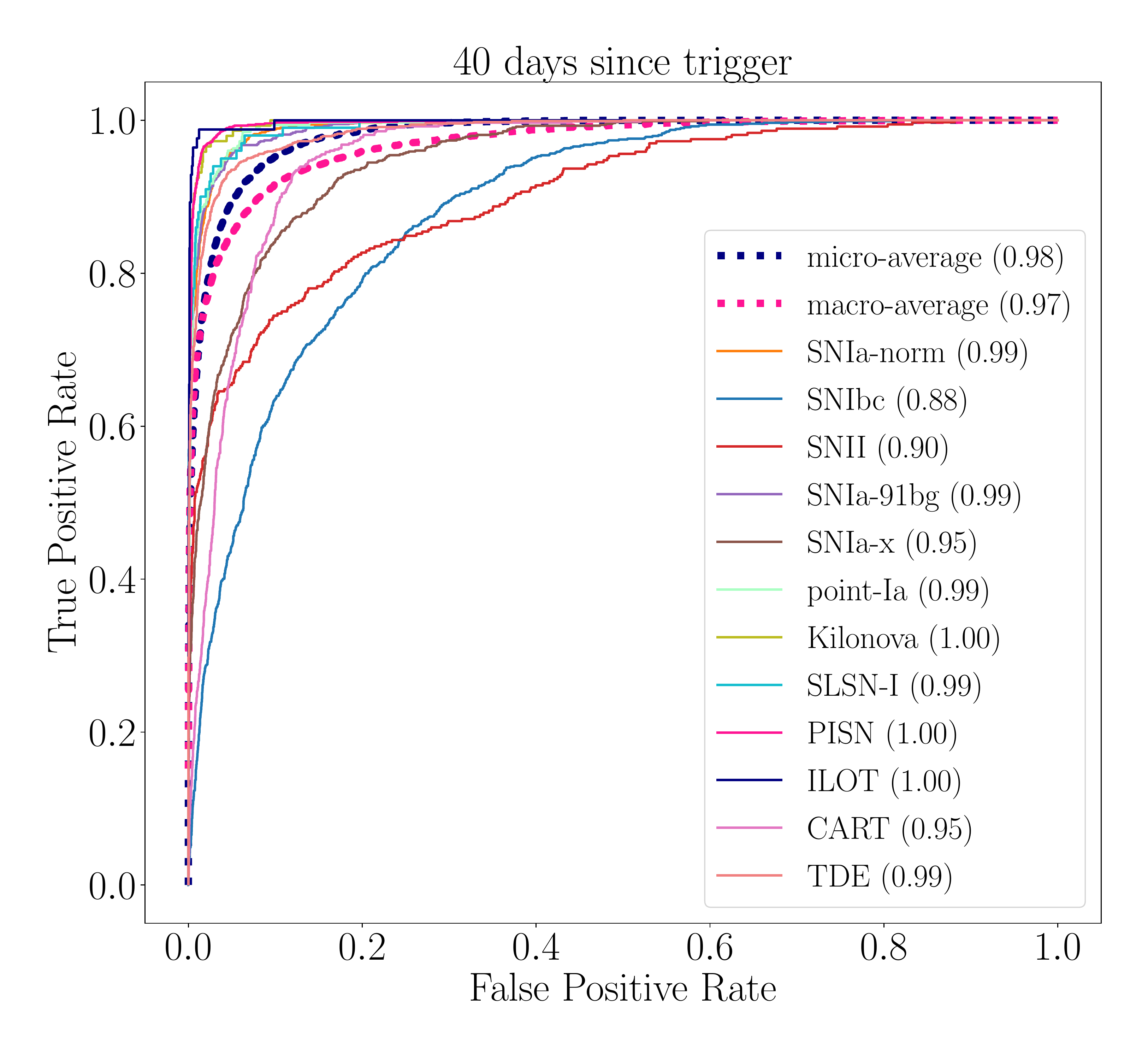}{0.5\textwidth}{(b) Late Epoch}
	}
	\caption{Receiver operating characteristic (ROC) curves for the 12 transient classes at an early epoch at 2 days past trigger (left), and at a late epoch at 40 days past trigger (right). Each curve represents a different transient class with the area under the ROC curve (AUC) score in the brackets. The macro-average and micro-average curves which are an average and weighted-average representation of all classes, respectively (see section \ref{sec:Performance}) are also plotted. We compute the metric on the 40\% of the dataset used for \textit{testing}. Please see the online material for an animation of the evolution of the ROC curve as a function of time since trigger. (\url{https://www.ast.cam.ac.uk/~djm241/rapid/roc.gif})}
	\label{fig:ROC}
    \end{figure*}

    The confusion matrix is a good measure of the performance of a classifier, but it does not make use of the full suite of probability vectors we obtain for every transient, and instead only uses the highest scoring class. The Receiver Operating Characteristic (ROC) Curve, on the other hand, makes use of the classification probabilities. Instead of selecting just the highest probability class, we use a probability threshold $p_{\text{thresh}}$. For each class $c$, transient $s$ at time $t$ is considered to be classified as $c$ if $y^{st}_c > p_{\text{thresh}}$. We sweep the values of $p_{\text{thresh}}$ between $0$ and $1$. The ROC curve plots the true positive rate (TPR) against the false positive rate (FPR) for these different probability thresholds. In a multi-class framework, the TPR is a measure of recall or completeness; it is the ratio between the number of correctly classified objects in a particular class (TP) to the total number of objects in that class (TP + FN). 
    \begin{equation}
        \mathrm{TPR = \frac{TP}{TP+FN}}
    \end{equation}
    Conversely, the FPR is a measure of the false alarm rate; it is the ratio between the number of transients that have been misclassified as a particular class (FP) and the total number of objects in all other classes (FP + TN).
    \begin{equation}
        \mathrm{FPR = \frac{FP}{FP+TN}}
    \end{equation}
    A good classifier is one that maximizes the area under the ROC curve (AUC), with a perfect classifier having an AUC=1, and a randomly guessing classifier having an AUC=0.5. Typically, values above 0.9 are considered to be very good classifiers. In Fig.~\ref{fig:ROC}, we plot the ROC curve at an early and late phase in the light curve. Here, the classification performance looks very good with several classes having AUC values above 0.99 and the overall micro-averaged values being 0.95 for the early stage and 0.98 in the late stage. The macro-averaged ROC is simply the average of all of the ROC curves computed independently. Differently, the micro-averaged ROC aggregates the TPR and FPR contributions of all classes, and is equivalent to the weighted average of all the ROCs considering the number of transients in each class. As the class distribution in the dataset is not too unbalanced, these values are quite close.
    
    \begin{figure}[htpb]
	\centering
	\includegraphics[width=1.\linewidth]{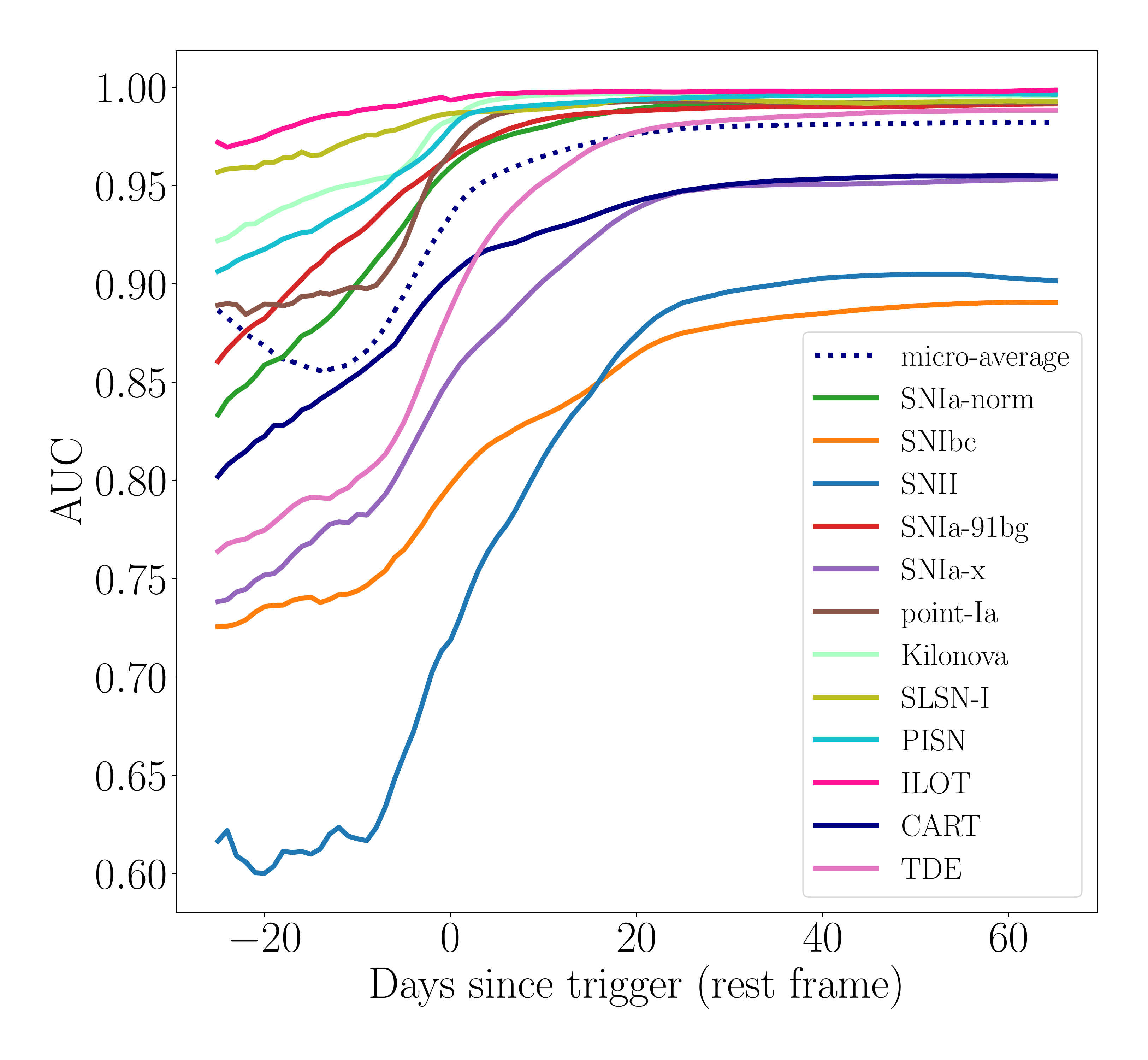}
	\caption{The area under the ROC curve (AUC) of each class as a function of time since trigger. Fig.~\ref{fig:ROC} illustrates the ROC curves at two epochs with the AUC of each listed in the legend; this plot shows hows the AUC evolves with time for each class, and is a still-representation of the animation of the ROC curves shown in the online material. The overall performance of the classifier is best judged with the shape of the `micro-average' curve.}
	\label{fig:auc_vs_time}
	\end{figure}
	
    \begin{figure*}
	\gridline{
		\fig{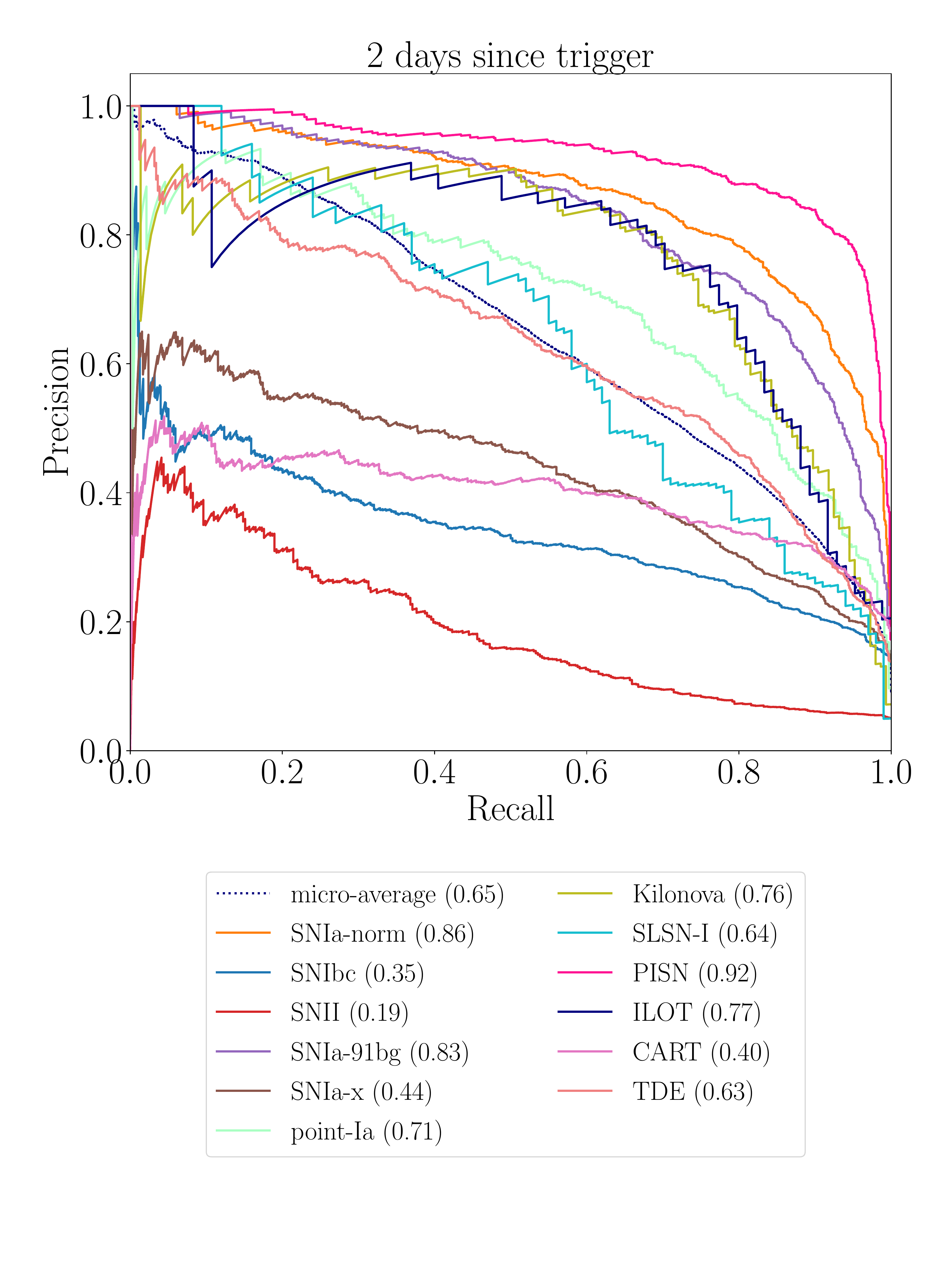}{0.5\textwidth}{(a) Early Epoch}
		\fig{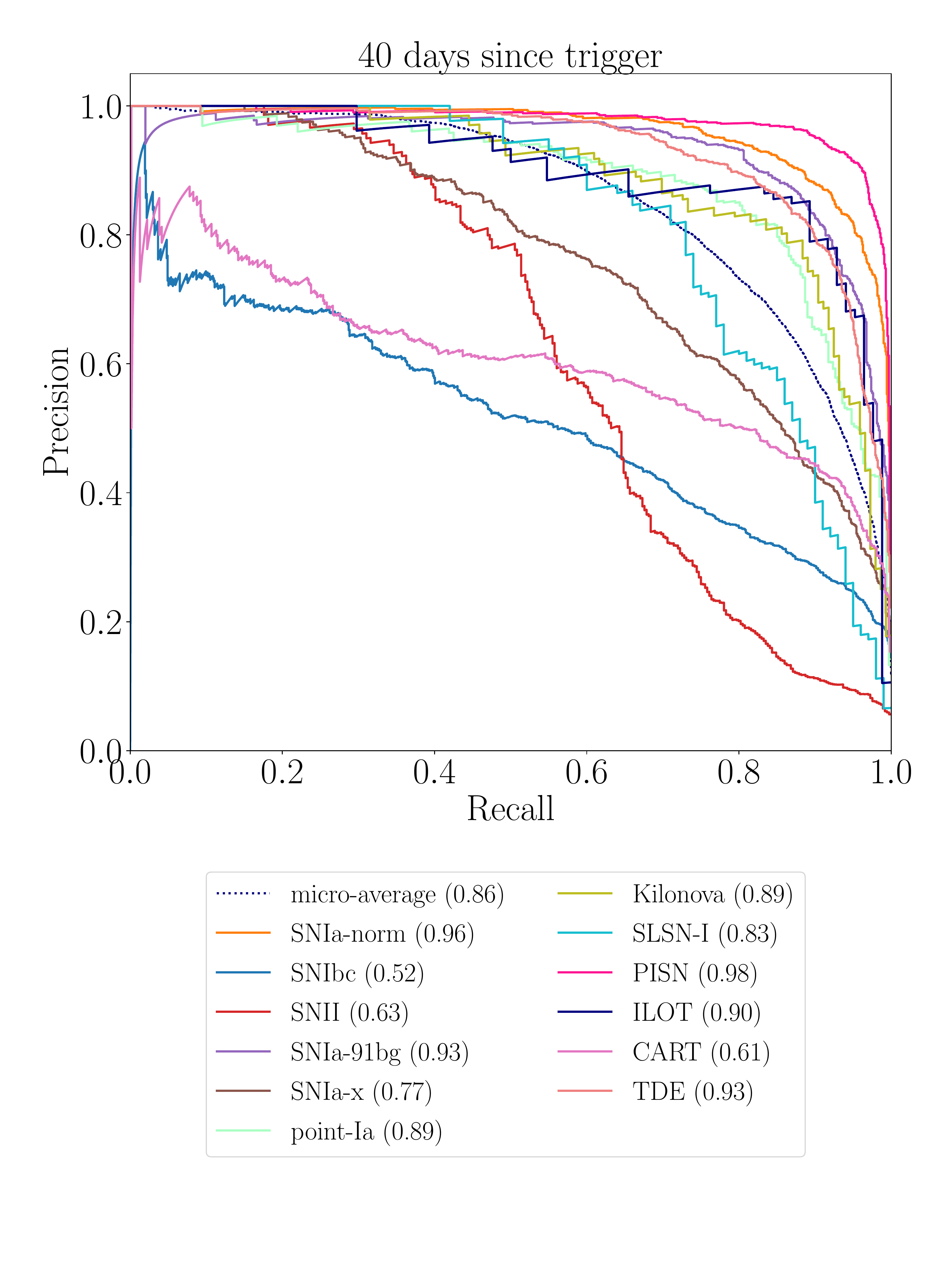}{0.5\textwidth}{(b) Late Epoch}
	}
	\caption{Precision-recall curves for the 12 transient classes at an early epoch at 2 days past trigger (left), and at a late epoch at 40 days past trigger (right). We compute the metric on the 40\% of the dataset used for \textit{testing}. Please see the online material for an animation of the evolution of the Precision-Recall as a function of time since trigger. (\url{https://www.ast.cam.ac.uk/~djm241/rapid/pr.gif}) }
	\label{fig:PR}
    \end{figure*}
    
    In the online material we plot an animation of the ROC curve evolving in time since trigger, rather than the two phases plotted here. As a still-image measure of this, we plot the AUC of each class as a function of time since trigger in Fig.~\ref{fig:auc_vs_time}. Within 5 to 20 days after trigger, the AUC flattens out for most classes. We see that the ILOTs, kilonovae, SLSNe, and PISNs are predicted with high accuracy well before trigger. These transients are fainter than most other classes, and hence, do not trigger an alert until their light curves approach maximum brightness. This means, that by the time a trigger happens, the transient behaviour of the light curve is mature, and the classifier has more information to be confident in its prediction. On the other hand, the two core collapse supernovae, SNII and SNIbc, have comparatively low AUCs. We expect that additional colour information will help to separate these from other supernova types. The overall performance is best illustrated by the micro-averaged AUC curve shown as the dotted blue curve. The AUC is initially low due to the misclassifications with Pre-explosion, but within just a few days after trigger, plateaus to a very respectable AUC of 0.98.
    
    \subsection{Precision-Recall}
    
    We compute the Precision-Recall metric. This metric has been shown to be particularly good for classifiers trained on imbalanced datasets \citep{Saito2015PRvsROC}. The precision (also known as purity) is a measure of the number of correct predictions in each class compared to the total number of predictions of that class, and is defined as,
    \begin{equation}
        \mathrm{precision = \frac{TP}{TP+FP}}.
    \end{equation}
    The Recall (also known as completeness) is the same as the true positive rate. It is a measure of the number of correct predictions in each class compared to the total number of that class in the testing set, and is defined as,
    \begin{equation}
        \mathrm{recall = \frac{TP}{TP+FN}}.
    \end{equation}
    A good classifier will have both high precision and high recall, and hence the area under the precision-recall curve will be high. In making the precision-recall plot, instead of simply selecting the class with the highest probability for each object, we apply a probability threshold as plotted in Fig.~\ref{fig:PR}. By using a very high probability threshold (instead of just selecting the most probable class), we can obtain a much more pure subset of classifications. The PISN, SNIa-norm, SNIa-91bg, kilonovae, pointIa, and ILOTs have very good precision and recall at the late epoch, and quite respectable at the early phase. The core collapse SNe and CARTs are again shown to perform poorly. Overall, this plot highlights some flaws in the classifier that the previous metrics did not capture. In particular, the CART class is shown to perform much more poorly than in previous metrics, highlighting that it does not have a high precision and that there are many false positives for it. As there are fewer CARTs in the test set than other classes, this was not as obvious in the other metrics (see \citet{Saito2015PRvsROC} for an analysis of precision-recall vs ROC curves as classification metrics).
    
    \subsection{Weighted Log Loss}
        \begin{figure}[htpb]
    	\centering
    	\includegraphics[width=1.\linewidth]{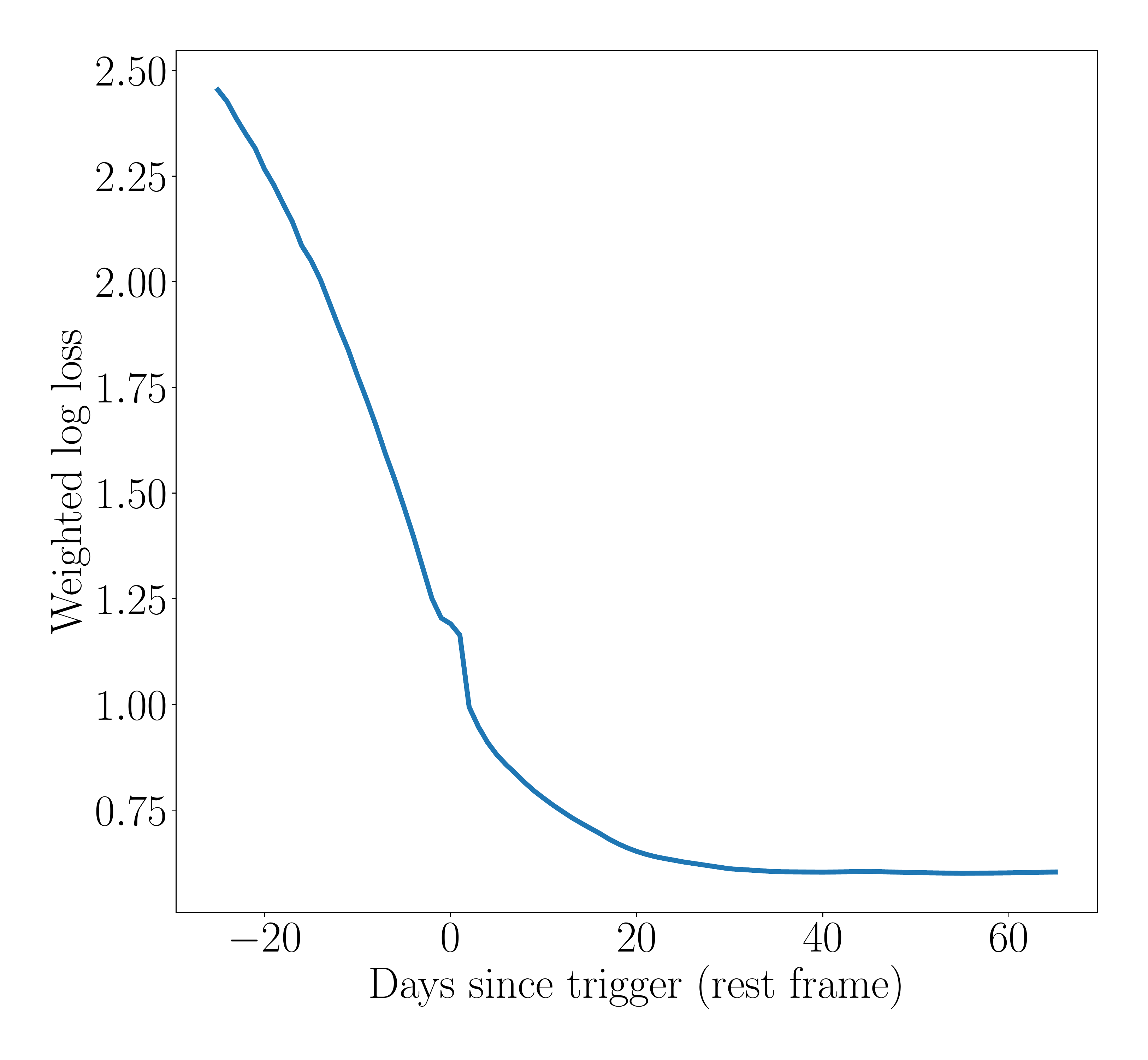}
    	\caption{The weighted log loss defined in equation \ref{eq:wlogloss} and used in PLAsTiCC \citep{plasticcMetric} is plotted as a function of time since trigger. The weighted log-loss of the early (2 days after trigger) and late (40 days after trigger) epochs are 1.09 and 0.64, respectively.}
    	\label{fig:wlogloss_vs_time}
	    \end{figure}
        In each of the previous metrics we have treated each class equally. However, it is often useful to weight the classifications of particularly classes more favourably than others. \citet{plasticcMetric} recently explored the sensitivity of a range of different metrics of classification probabilities under various weighting schemes. They concluded that a weighted log-loss provided the most meaningful interpretation, defined as follows
        \begin{equation}
            \ln \mathrm{Loss}^t= - \Bigg( \frac{\sum_{c=1}^{(m+1)} w_c \cdot \sum_{s=1}^{N_c} \frac{Y^{st}_{c}}{N_c} \cdot \ln y^{st}_c} {\sum_{c=1}^{(m+1)} w_c} \Bigg)
        \label{eq:wlogloss}
        \end{equation}
        where $c$ is an index over the $(m+1)$ classes and $j$ is an index over all $N$ members of each class, $y^{st}_{c}$ is the predicted probability that object $s$ at time $t$ is a member of class $c$, and $\bm{Y}^{st}$ is the truth label. The weight of each class $w_c$ can be different, and $N_c$ is the number of objects in each class.

        This metric is currently being used in the PLAsTiCC Kaggle challenge to assess the classification performance of each entry. We apply a weight of $2$ to the classes that PLAsTiCC deemed to be rare or interesting and $1$ to the remaining classes: 
        \begin{description}
            \item[Weight 1] SNIa, SNIbc, SNII, Ia-91bg, Ia-x, Pre-explosion
            \item[Weight 2] Kilonova, SLSN, PISN, ILOT, CART, TDE
        \end{description}
        We plot the weighted log-loss as a function of time since trigger in Fig.~\ref{fig:wlogloss_vs_time}. The metric of the early (2 days after trigger) and late (40 days after trigger) epochs are 1.09 and 0.64, respectively, where a perfect classifier receives a score of 0. While we have applied our classifier to ZTF simulations, we find that the raw scores are competitive with top scores in the PLAsTiCC Kaggle challenge. The sharp improvement in performance at trigger is primarily due to the prior placed on the Pre-explosion phase of the light curve that forces it to be before trigger. Within approximately 20 days after trigger, the classification performance plateaus, as the transient phase of most light curves is ending.
    
\section{Application to Observational Data}
\label{sec:application_to_real_data}
    One of the primary challenges with developing classifiers for astronomical surveys is obtaining a labelled sample of well-observed transients across a wide range of classes. While it may be possible to obtain a labelled sample of common supernovae during the initial stages of a survey, the observation rates of less common transients (such as kilonovae and CARTs, for example) mean that a diverse and large training set of observed data is impossible to obtain. Therefore, a classifier that is trained on simulated data but can classify observational data streams is of significant importance to the astronomical community. To this end, a key goal of \texttt{RAPID} is to be able to classify observed data using an architecture trained on only simulated light curves. In this section, we provide a few examples of \texttt{RAPID}'s performance on transients from the ongoing ZTF data stream. In future work, we hope to extend this analysis to test the classification performance on a much larger set of observed light curves.

    \begin{figure*}[htpb]
	\centering
	\includegraphics[width=1.\linewidth]{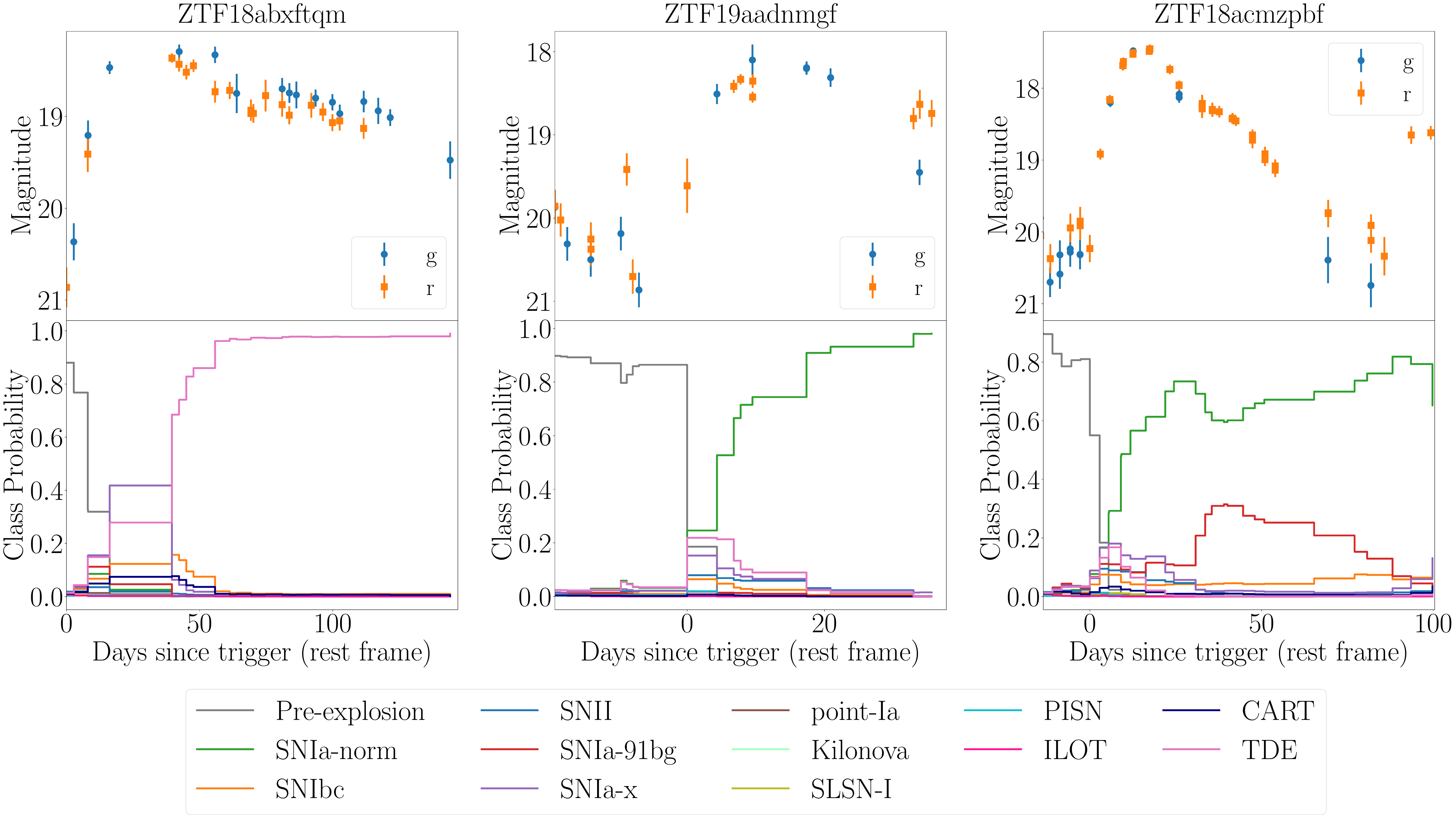}
	\caption{Classification of three light curves from the observed ZTF data stream. In each of the three example cases, \texttt{RAPID} correctly classifies the transient well before peak brightness, and often within just a few epochs. The ZTF names are listed in the titles of each transient plot and from left to right they are also known by the following names: AT2018hco, SN2019bly, SN2018itl. These objects were spectroscopically classified as a TDE ($z=0.09$), SNIa ($z=0.08$), and SNIa ($z=0.036$), respectively \citep{ATelAT2019hcoVelzen,ATelSN2019blyFremling,ATelSN2018itlFremling}.}. 
	\label{fig:real_ZTF}
	\end{figure*}

    In Fig.~\ref{fig:real_ZTF} we have tested \texttt{RAPID} on three objects recently observed by ZTF to highlight its direct use on observational data: ZTF18abxftqm, ZTF19aadnmgf, and ZTF18acmzpbf (also known as AT2018hco\footnote{\url{https://wis-tns.weizmann.ac.il/object/2018hco}}, SN2019bly\footnote{\url{https://wis-tns.weizmann.ac.il/object/2019bly}}, SN2018itl\footnote{\url{https://wis-tns.weizmann.ac.il/object/2018itl}}, respectively). These have already been spectroscopically classified as a TDE ($z=0.09$), SNIa ($z=0.08$), and SNIa ($z=0.036$), respectively \citep{ATelAT2019hcoVelzen,ATelSN2019blyFremling,ATelSN2018itlFremling}. In the bottom panel of Fig.~\ref{fig:real_ZTF}, we see that \texttt{RAPID} was able to correctly confirm the class of each transient well before maximum brightness, and within just a couple of epochs after trigger.
    
    The two SNIa light curves were correctly identified after just one detection epoch, and the confidence in these classifications improved over the lifetime of the transient. While the SNIa-norm probability was lower for ZTF18acmzpbf, this is a good example of where the confidence in this transient being any subtype of SNIa was actually much higher. Given that the second most probable class was a SNIa-91bg, we can sum the two class probabilities to obtain a much higher probability of the transient being a SNIa.
    
    While we have shown \texttt{RAPID}'s effective performance on some observational data, future revisions of the software can be used to identify differences between the simulated training set and observations. This will help to improve the transient class models that were used to generate the light curve simulations. Future iterations to improve the simulations will in turn lead to a classifier that is even more effective at classifying observational data.
    
    Moreover, as it stands, \texttt{RAPID} can classify 12 different transient classes. However, if an unforeseen transient were passed into the classifier, the class probabilities would split between the classes that were most similar to the input. \texttt{RAPID} is a supervised learning algorithm, and is not designed for anomaly detection. However, cases where \texttt{RAPID} is not confident on a classification may warrant closer attention for the possibility of an unusual transient.

    \subsection{Balanced or representative datasets}
        Machine learning based classifiers such as neural networks often fail to cope with imbalanced training sets as they are sensitive to the proportions of the different classes \citep{kotsiantis2006ImbalancedData}. As a consequence, these algorithms tend to favour the class with the largest proportion of observations. This is particularly problematic when trying to classify rare classes. The intrinsic rate of some majority classes, such as SNe Ia, is orders of magnitude higher than some rare classes, such as kilonovae. A neural network classifier trained on such a representative dataset, and that aims to minimise the overall unweighted objective function (equation \ref{eq:objective}), will be incentivised to learn how to identify the most common classes rather than the rare ones. An example of the poorer performance of classifiers trained on representative transient datasets compared to balanced datasets is illustrated well by Figure 7 in \citet{Narayan2018MachineStream}. The shown t-SNE (t-distributed Stochastic Neighbour Embedding, \citealt{tSNE}) plot is able to correctly cluster classes much more accurately when the dataset is balanced.
        
        Moreover, building a representative dataset is a very difficult task and there is a non-representativeness between spectroscopic and photometric samples. The detection rates of different transient classes in photometric surveys are often biased by brightness and the ease at which some classes can be classified over others. Spectroscopic follow-up strategies have been dominated by SNe Ia for cosmology, and have hence led to biased spectroscopic sets. Recently, however, \citet{Ishida2019} identified a framework for constructing a more representative training set. They make use of real-time active learning to improve the way labelled samples are obtained from spectroscopic surveys to ultimately optimise the scientific performance of photometric classifiers. Employing such a framework will allow for the construction of a training set that is more representative. 
        
        In our work, we simulate a balanced training set in an attempt to mitigate the effects of the bias present in existing datasets and to improve our classifier's accuracy on rare classes. While machine learning classifiers tend to perform better on balanced datasets \citep{kotsiantis2006ImbalancedData,Narayan2018MachineStream}, future work should verify this for photometric identification by comparing the performance of classifiers trained on balanced and representative training sets. However, until the issue of the non-representativeness between spectroscopic and photometric samples is mitigated with approaches like \citet{Ishida2019}, a representative dataset remains difficult to build.

\section{Feature-based Early Classification}
    \label{sec:RandomForestClassifier}
    
    To compare the performance of \texttt{RAPID} against traditional light curve classification approaches which often use extracted light curve features for classification, we developed a Random Forest-based classifier that computed statistical features of each light curve to use as input, rather than directly using the photometric information. This has been the most commonly used approach in light curve classification tasks to date \citep[e.g.][]{Lochner2016,Narayan2018MachineStream, Moller2016,Newling2011,Karpenka2012ANetworks,Revsbech2017STACCATO:Sets}. We extend upon the approach developed in \citet{Narayan2018MachineStream} and based on \citet{Lochner2016} by using a wider variety of important features and by extending the problem for early light curve classification. Specifically, we only compute features using data up to 2 days after trigger so that the Random Forest classifier can be directly compared with the DNN early classifications. 

    Extracting features from light curves provides us with a uniform method of comparing between transients which are often unevenly sampled in time. We can train directly on a synthesised feature vector instead of the photometric light curves. For time-series data, extracting moments is the most obvious way to start obtaining features. We compute several moments of the light curve, and a list of the distilled features used in classification are listed in Table \ref{tab:feature_table}. While we focus on early classification in this paper, we also list some full-light curve features that we used in work not shown in this paper that some readers may find useful. As we have two different passbands, we compute the features for each passband and obtain twice the number of moment-based features listed in the table. We also make use of more context specific features, such as redshift and colour information.

        \begin{table*}
            \centering
            \begin{tabular}{|m{4.5cm}|m{10.5cm}|}
                 
                 \hline \hline
                 \multicolumn{2}{|c|}{\textbf{Early light curve features only}} \\
                 \hline
                 Early rise rate &  Slope of early light curve (see equation \ref{eq:riserate}). \\
                 \hline
                 $a$ &  Amplitude of quadratic fit to the early light curve (see equation \ref{eq:model_early}). \\
                 \hline
                 $c$ &  Intercept of quadratic fit to the early light curve (see equation \ref{eq:model_early}). \\
                 \hline
                 Colour at n days &  Logarithmic ratio of the flux in two passbands (see equation \ref{eq:colour}). \\
                 \hline
                 Early colour slope &  Slope of the colour curve. \\
                 
                 \hline \hline
                 \multicolumn{2}{|c|}{\textbf{Early and full light curve features}} \\
                 \hline
                 Redshift &  Photometric cosmological redshift. \\
                 \hline
                 Milky Way Dust Extinction &  Interstellar extinction. \\
                 \hline
                 Historic Colour &  Logarithmic ratio of the flux in two passbands before trigger. \ref{eq:colour}).   \\
                 \hline
                 Variance &  Statistical variance of the flux distribution.\\
                 \hline
                 Amplitude &  Ratio of the 99th minus 1st and 50th minus 1st percentile of the flux distribution. \\
                 \hline
                 Standard Deviation / Mean &  A measure of the average inverse signal-to-noise ratio. \\
                 \hline
                 Median Absolute Deviation &  A robust estimator of the standard deviation of the distribution. \\
                 \hline
                 Autocorrelation Integral &  The integral of the correlation vs time difference \citep{Mislis2015SIDRA:Surveys}. \\
                 \hline
                 Von-Neumann Ratio &  A meausure of the autocorrelation of the flux distribution. \\
                 \hline
                 Entropy &  The Shannon entropy assuming a Gaussian CDF following \citet{Mislis2015SIDRA:Surveys}. \\
                 \hline
                 Rise time &  Time from trigger to peak flux. \\
                 
                 \hline \hline
                 \multicolumn{2}{|c|}{\textbf{Full light curve features only}} \\
                 \hline
                 Kurtosis &  Characteristic ``peakedness'' of the magnitude distribution.   \\
                 \hline
                 Shapiro-Wilk Statistic &  A measure of the flux distribution's normality. \\
                 \hline
                 Skewness &  Characteristic asymmetry of the flux distribution. \\
                 \hline
                 Interquartile Range &  The difference between the 75th and 25th percentile of the flux distribution. \\
                 \hline
                 Stetson K &  An uncertainty weighted estimate of the kurtosis following \citet{Stetson1996OnVariables}. \\
                 \hline
                 Stetson J &  An uncertainty weighted estimate of the Welch-Stetson Variability Index \citep{Welch1993Robust1866}. \\
                 \hline
                 Stetson L &  Product of the Stetson J and Stetson K moments \citep{Kinemuchi2006AnalysisSurvey,Stetson1996OnVariables}. \\
                 \hline
                 HL Ratio &  The ratio of the amplitudes of points higher and lower than the mean. \\
                 \hline
                 Fraction of observations above trigger &  Fraction of light curve observations above the trigger. \\
                 \hline
                 Period &  Top ranked periods from the Lomb-scargle periodogram fit of the light curves \citep{Lomb1976Least-squaresData,Scargle1982StudiesData}. \\
                 \hline
                 Period Score &  Period weights from the Lomb-scargle periodogram fit of the light curves \citep{Lomb1976Least-squaresData,Scargle1982StudiesData}. \\
                 \hline
                 Colour Amplitude & Ratio of the amplitudes in two passbands. \\
                 \hline
                 Colour Mean & Ratio of the mean fluxes in two passbands. \\
                 \hline
                 
            \end{tabular}
            \caption{Description of the features extracted from each passband of each light curve in the dataset. Some of these are redefined from Table 2 of \citet{Narayan2018MachineStream}.}
            \label{tab:feature_table}
        \end{table*}
        \normalsize
    
    We compute the \textit{early rise rate} feature for each passband as the slope of the fitted early light curve model defined in section \ref{sec:tsquare},
    \begin{equation}
        \mathrm{rate}^\lambda = \frac{L_\mathrm{mod}^\lambda(t_{\mathrm{peak}}) - {L_\mathrm{mod}^\lambda(t_{0})}}{(t_{\mathrm{peak}} - t_0)}.
        \label{eq:riserate}
    \end{equation}
	We use the rise rate, and the early light curve model parameter fits $\bm{\hat{a}}$ and $\bm{\hat{c}}$ from equation \ref{eq:model_early} as features in the early classifier. We then define the colour as a function of time,
    \begin{equation}
        \mathrm{colour}(t) = -2.5 \log_{10}{\left( \frac{L_\mathrm{mod}^{g}(t)}{L_\mathrm{mod}^{r}(t)} \right)},
        \label{eq:colour}
    \end{equation}
    where $L_\mathrm{mod}^{\lambda}(t)$ is the modelled relative luminosity (defined in equation \ref{eq:model_early}) at a particular passband, $\lambda$.

	We use the colour curves computed from each transient to define several features. Using equation \ref{eq:colour}, we compute the colour of each object at a couple of well-spaced points on the early light curve (5 days and 9 days after $t_0$) and use them as features in our early classifier. We also compute the slope of the colour curve and use that as an additional feature for the early classifier. For the full light curve classifier, we compute the colour amplitude as the difference in the light curve amplitudes in two different passbands, and also compute the colour mean as the ratio of the mean flux value of two different passbands.
        
		\begin{figure}
    	\centering
    	\includegraphics[width=1.\linewidth]{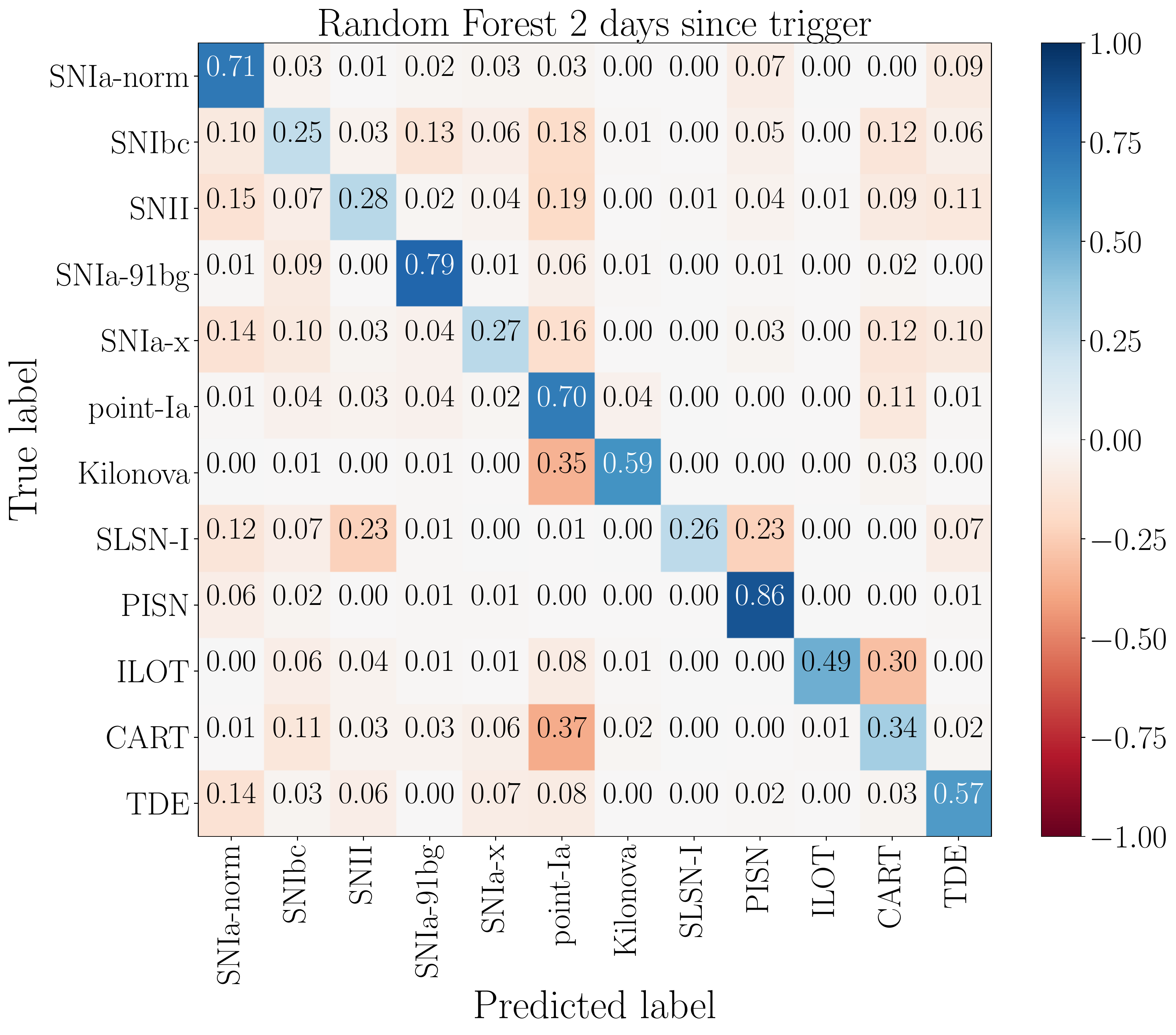}
    	\caption{Confusion matrix of the early light curve Random Forest classifier trained on 60\% and tested on 40\% of the dataset described in section \ref{sec:Data}. The classifier makes use of 200 estimators (trees) in the ensemble. The colour bar and values indicate the percentage of each \textit{true label} that were classified as the \textit{predicted label}. Negative colour bar values are used only to indicate misclassifications.}
    	\label{fig:confusion_matrix_RandomForest_early}
    	\end{figure}
    	
    	\begin{figure}
    	\centering
    	\includegraphics[width=1.\linewidth]{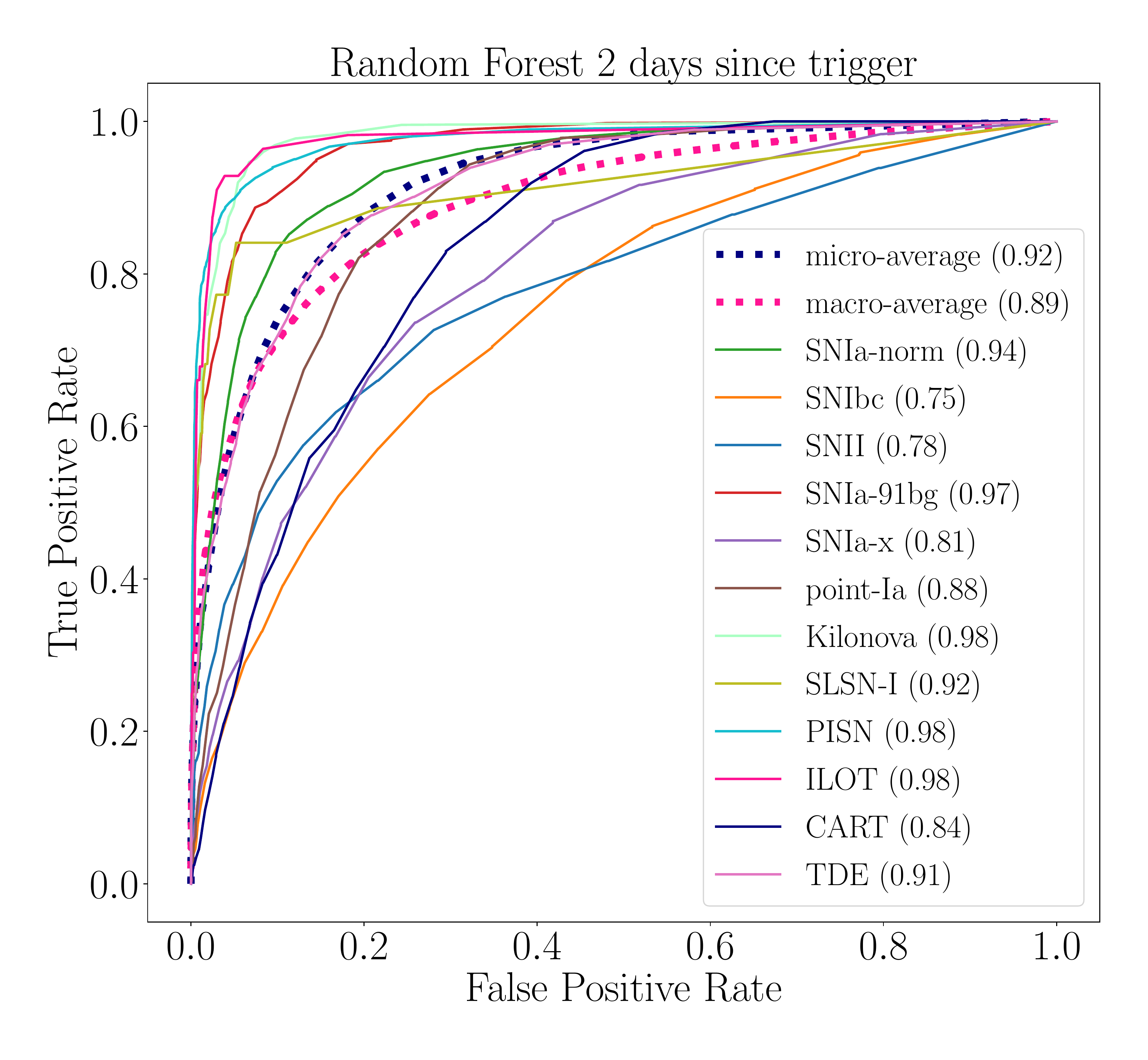}
    	\caption{Receiver operating characteristic of the feature-based Random Forest approach. The features are computed on photometric data up to 2 days past trigger, and are fed into a Random Forest classifier. Each curve represents a different transient class with the area under the curve (AUC) score in brackets. The macro and micro average curves which are an average and weighted-average representation of all classes are also plotted. The metric computed on 40\% of the dataset.}
    	\label{fig:roc_RandomForest_early}
    	\end{figure}
    	
    	\begin{figure*}
    	\centering
    	\includegraphics[width=1.\linewidth]{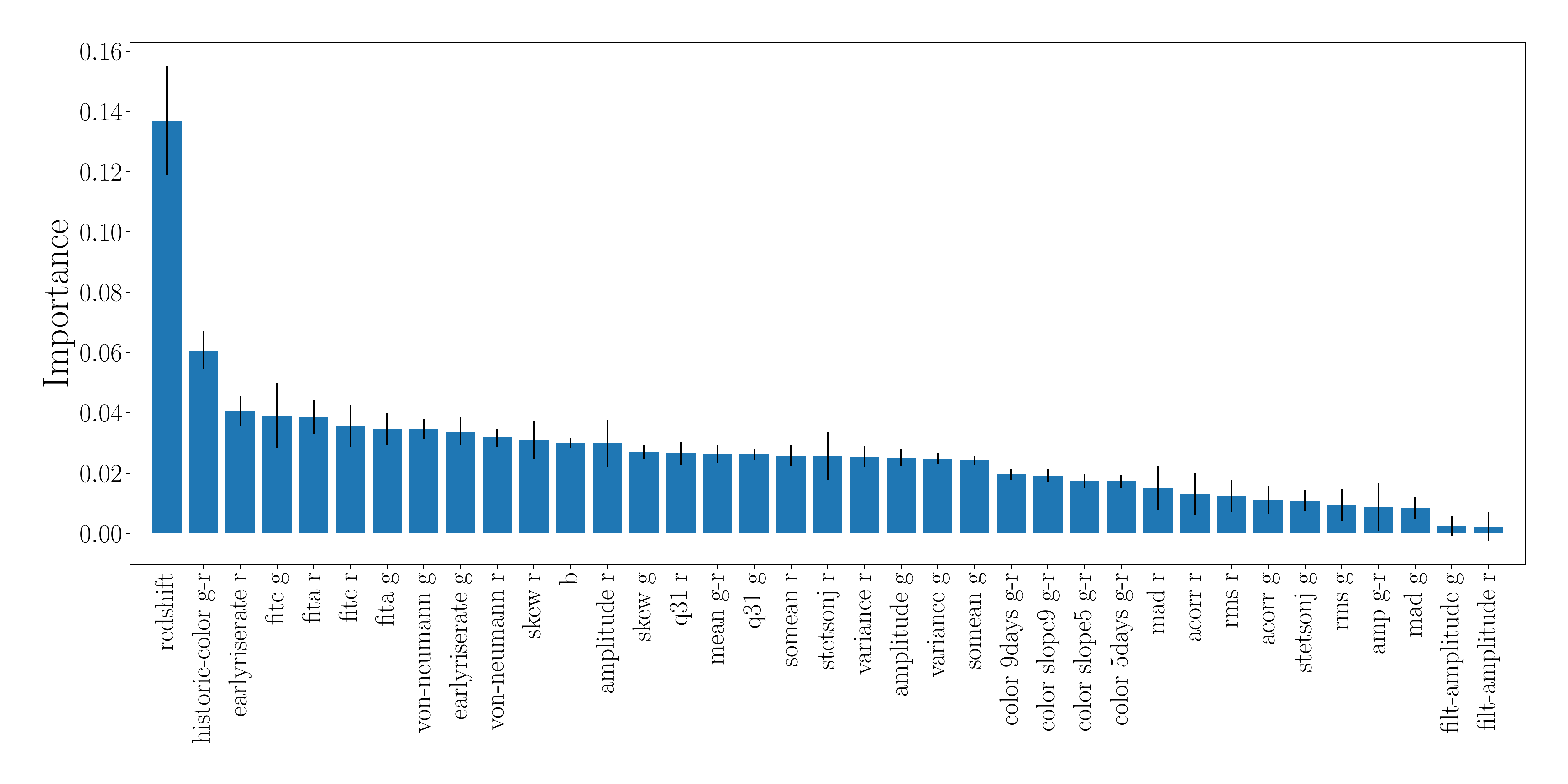}
    	\caption{Features used in the early classifier ranked by importance. The bar height is the average relative importance of each feature in each tree of the random forest. The error bar is the standard deviation of the importance of each feature in the 200 trees of the Random Forest. Only the top 30 features have been plotted in the histogram.}
    	\label{fig:feature_importance_early}
    	\end{figure*}
    	
	    We feed the feature set into a Random Forest classifier. Random Forests \citep{Breiman2001RandomForests} are one of the most flexible and popular machine learning architectures. They construct an ensemble of several fully grown and uncorrelated decision trees \citep{Morgan1963ProblemsProposal} to create a more robust classifier that limits overfitting. Each decision tree is made up of a series of hierarchical \textit{branches} that check whether values in the feature vector are in a particular range until it ascertains each of the class labels in the form of \textit{leaves}. The trees are trained recursively and independently, selecting which feature and boundary provide the highest information gain for classifications. A single tree is subject to high variance and can easily overfit the training set. By combining an ensemble of decision trees - providing each tree with a subset of data that is randomly replaced during training - a Random Forest is able to decrease the variance by averaging the results from each tree.
    	
    	We have designed the Random Forest with 200 estimators (or trees) and have run it through twice. On the first run we feed the classifier the entire feature-set. We then rank the features by importance in classification and select the top 30 features. We feed only these top 30 features into the second run of the classifier. As many of the features are obviously highly correlated with each other, this acts to reduce \textit{feature dilution}, whereby we remove features that do not provide high selective power.
    	
        We compute features using light curves up to only the first $2$ days after trigger. As the Random Forest is much quicker to classify than the DNN, we perform 10-fold cross-validation to obtain a more robust estimate of the classifier's performance. We then produce the confusion matrix in Fig.~\ref{fig:confusion_matrix_RandomForest_early} and the ROC curve in Fig.~\ref{fig:roc_RandomForest_early}. We can compare these metrics to the early epoch metrics at $2$ days after trigger produced with the deep neural network in Figures \ref{fig:RNN_CF} and \ref{fig:ROC}. We find that the performance in the early part of the light curve is marginally worse than the DNN with a micro-averaged AUC of 0.92 compared to 0.95. Moreover, the ability of the DNN to provide time-varying classifications makes it much more suited to early classification than the Random Forest.
        
    	In Fig.~\ref{fig:feature_importance_early}, we rank the importance of the top $30$ features in the Random Forest classifier. While redshift is clearly the most important feature in the dataset, we have also built classifiers without using redshift as a feature and found that the performance was only marginally worse. This provides an insight into the classifier's robustness when applied to surveys where redshift is not available. The next best features is the historic colour, suggesting that the type of host-galaxy is important contextual information to discern transients. The early slope of the light curve also ranks highly, as it is able to distinguish between faster rising core-collapse supernovae and other slower rising transients such as magnetars (SLSNe).
    	
\section{Conclusions}
    \label{sec:Conclusions}
    
    Existing and future wide-field optical surveys will probe new regimes in the time-domain, and find new astrophysical classes, while enabling a deeper understanding of presently rare classes. In addition, correlating these sources with alerts from gravitational wave, high-energy particle, and neutrino observatories will enable new breakthroughs in multi-messenger astrophysics. However, the alert-rate from these surveys far outstrips the follow-up capacity of the entire astronomical community combined. Realising the promise of these wide-field surveys requires that we characterize sources from sparse early-time data, in order to select the most interesting objects for more detailed analysis. 
    
    We have detailed the development of a new real-time photometric classifier, \texttt{RAPID}, that is well-suited for the millions of alerts per night that ongoing and upcoming wide-field surveys such as ZTF and LSST will produce. The key advantages that distinguish our approach from others in the literature are:
    \begin{enumerate}
        \item Our deep recurrent neural network with uni-directional gated recurrent units, allows us to classify transients using the available data as a function of time.
        \item Our architecture combined with a diverse training set allows us to identify 12 different transient classes \emph{within days of its explosion, despite low S/N data and limited colour information}.
        \item We do not require user-defined feature extraction before classification, and instead use the processed light curves as direct inputs. 
        \item Our algorithm is designed from the outset with speed as a consideration, and it can classify the tens of thousands of events that will be discovered in each LSST image within a few seconds. 
    \end{enumerate}
    
    This critical component of \texttt{RAPID} that enables early classification is our ability to use measurements of the source before an alert is triggered --- ``precovery'' photometry with insufficient significance to trigger an alert, but that nevertheless encodes information about the transient. While we designed \texttt{RAPID} primarily for early classification, the flexibility of our architecture means that it is also useful for photometric classification with any available phase coverage of the light curves. It is competitive with contemporary approaches such as \citet{Lochner2016,Charnock2016,Narayan2018MachineStream} when classifying the full light curve. 
    
    There is no satisfactory single metric that can completely summarise classifier performance, and we have presented detailed confusion matrices, ROC curves and measures of precision vs recall for all the classes represented in our training set. The micro-averaged AUC, the most common single metric used to measure classifier performance, evaluated across the 12 transient classes is 0.95 and 0.98 at 2 days and 40 days after an alert trigger, respectively. We further evaluated \texttt{RAPID}'s performance on a few transients from the real-time ZTF data stream, and, as an example, have shown its ability to effectively identify a TDE and two SNe Ia well before maximum brightness. The results at early-times are particularly significant as, in many cases, they can exceed the performance of trained astronomers attempting visual classification.
    
    We also developed a second early classification approach that trained a Random Forest classifier on features extracted from the light curve. This allowed us to directly compare the feature-based Random Forest approach to \texttt{RAPID}'s model-independent approach. We found that the classification performances are comparable, but the RNN has the advantage of obtaining time-varying classifications, making it ideal for transient alert brokers. To this end, we have recently begun integrating the \texttt{RAPID} software with the ANTARES alert-broker, and plan to apply our DNN to the real-time ZTF data stream in the near future. 
    
    In future work, we plan on applying this method on LSST simulations to help to inform how changes in observing strategy affect transient classifications at early and late phases. Overall, \texttt{RAPID} provides a novel and effective method of classifying transients and providing prioritised follow-up candidates for the new era of large scale transient surveys.
    
\acknowledgements
DM would like to thank the Cambridge Australia Poynton Scholarship and Cambridge Trust for studentship funding, as well as Armin Rest and the Director's Office at STScI for travel funding. GN is supported by the Lasker Fellowship at the Space Telescope Science Institute, and thanks the Kavli Institute for Cosmology, Cambridge for travel funding through a Kavli Visitor Grant. This work used the science cluster at STScI and the Midway cluster at the University of Chicago. We are indebted to IT staff at STScI, and the University of Chicago Research Computing Center for their assistance with our work. We thank Eric Bellm for providing a database of ZTF observations thank Rick Kessler for help with the SNANA simulation code used for making the ZTF simulated light curves, and finally thank Ryan Foley, David Jones and Zack Akil for useful conversations.

\software{{\tt AstroPy} \citep{Robitaille2013},
    {\tt Keras} \citep{Keras},
    {\tt NumPy} \citep{vanderWalt2011},
    {\tt Scikit-learn} \citep{scikit-learn},
    {\tt SciPy} \citep{Jones2001},
	{\tt TensorFlow} \citep{Abadi2015}.
}


\bibliographystyle{aasjournal}
\bibliography{references,ref2} 

\end{document}